\documentclass{emulateapj}
\usepackage{graphics,wrapfig,amsmath,color, float}% Include figure files
\begin{document}

\title{Characterization of Seven Ultra-Wide Trans-Neptunian Binaries}
\author{Alex H. Parker$^{1,2}$, JJ Kavelaars$^{3}$, Jean-Marc Petit$^{4}$, Lynne Jones$^{5}$, Brett Gladman$^{6}$, Joel Parker$^{7}$}
\email{alexhp@uvic.ca}
\affil{{ $^{1}$Department of Astronomy, University of Victoria, Victoria, BC, Canada\\
$^{2}$Institute for Theory and Computation, Harvard Smithsonian Center for Astrophysics, Cambridge, MA, USA\\
$^{3}$Herzberg Institute of Astrophysics, National Research Council of Canada, Saanich, BC, Canada\\
$^{4}$Observatoire de Besan\c{c}on, Besan\c{c}on, France\\
$^{5}$Department of Astronomy, University of Washington, Seattle, WA, USA\\
$^{6}$Department of Astronomy, University of British Columbia, Vancouver, BC, Canada\\
$^{7}$Southwest Research Institute, Boulder, CO, USA}}

\shortauthors{Parker et al.}

\begin{abstract}

The low-inclination component of the Classical Kuiper Belt is host to a population of extremely widely-separated binaries. These systems are similar to other Trans-Neptunian binaries (TNBs) in that the primary and secondary components of each system are of roughly equal size. We have performed an astrometric monitoring campaign of a sample of seven wide-separation, long-period TNBs and present the first-ever well-characterized mutual orbits for each system. The sample contains the most eccentric (2006 CH$_{69}$, $e_{m}=0.9$) and the most widely-separated, weakly bound (2001 QW$_{322}$, $a/R_H \simeq 0.22$) binary minor planets known, and also contains the system with lowest-measured mass of any TNB (2000 CF$_{105}$, $M_{sys} \simeq 1.85\times10^{17}$ kg). Four systems orbit in a prograde sense, and three in a retrograde sense. They have a different mutual inclination distribution compared to all other TNBs, preferring low mutual-inclination orbits. These systems have geometric $r$-band albedos in the range of $0.09-0.3$, consistent with radiometric albedo estimates for larger solitary low-inclination Classical Kuiper Belt objects, and we limit the plausible distribution of albedos in this region of the Kuiper Belt. We find that gravitational collapse binary formation models produce a similar orbital distribution to that currently observed, which along with a confluence of other factors supports formation of the cold Classical Kuiper Belt \textit{in situ} through relatively rapid gravitational collapse rather than slow hierarchical accretion. We show that these binary systems are sensitive to disruption via collisions, and their existence suggests that the size distribution of TNOs at small sizes remains relatively shallow.

\end{abstract}


\maketitle

%\clearpage

\section{Introduction}

Most, if not all, minor planet populations are host to multiple systems, including binaries and trinaries (see Walsh 2009 for review). These multiples have a vast array of properties, from extremely short-period, contact binaries, to systems that are exceedingly widely separated and have mutual periods of many years. Some have tiny satellites, thought to be collisional fragments blown off of their parent, and others have components of near-equal size and mass. The characteristics of these systems represent a treasure trove of information about the properties of the objects that compose them, the environment they are embedded in, and the dynamical history of their parent population. Through their orbital separations and periods, binaries offer the only way to measure the mass of these distant objects, which when combined with radius measurements determine these objects' bulk densities --- which in turn provide information about composition and physical structure (such as porosity).

In the last 10 years, Trans-Neptunian populations have been found to host to a very high fraction of binary systems. The binary fraction varies in sub-populations from $\sim30$\% in the low-inclination component of the Classical Kuiper Belt to just a few percent in other dynamical classes (Noll et al. 2008a). Given the low interaction rates of the Kuiper Belt populations today, forming such a large number of binary systems has proven a theoretical challenge, especially with the limited information available for the components of these systems (eg., Goldreich et al. 2002, Weidenschilling 2002, Noll et al. 2008b, Schlichting \& Sari 2008a \& b).

Trans-Neptunian Binaries (TNBs) are distinguished from binary systems elsewhere in the solar system by the high frequency of near-equal sized binaries, and by the presence of binaries with extremely wide separations and long mutual-orbit periods. Widely-separated, long-period TNBs are difficult to create and very sensitive to perturbation (Nesvorn\'{y} et al. 2011, Parker \& Kavelaars 2010, Petit \& Mousis 2004), and make valuable tracers of the dynamical and collisional conditions over the history of the outer solar system. The orbital, compositional and statistical properties of these binaries constrain the total mass and dynamical history of the various populations, with important implications for theories of Solar System formation and evolution.

There are several dozen known TNBs, but only a small subset has measured orbital parameters (Noll et al. 2008b, Naoz et al. 2010, Grundy et al. 2011). Most of these, in turn, are relatively tightly-bound binaries that have been characterized by observations from space (systems with mutual semi-major axis much less than 5\% of their Hill radius, eg., Grundy et al. 2009 \& 2011). Two TNB systems with moderately widely-separated components have published mutual orbits (1998 WW$_{31}$ and Teharonhiawako/Sawiskera, both with mutual semi-major axes of order 5\% of their Hill radius), but the widest TNBs (those with mutual semi-major axes substantially exceeding 5\% of their Hill sphere) have not been well-characterized to date, with a preliminary orbit estimate available only for the system 2001 QW$_{322}$ (Petit et al. 2008). Such wide-separation, near-equal mass binaries all have low heliocentric inclinations, indicating that they belong to the cold component of the Classical Kuiper Belt. These wide binaries make up at least 1.5\% of the known Cold Classical belt objects (Lin et al. 2010).

The creation of large-separation, near-equal mass binaries is most likely during the formation phase of TNOs, as most proposed formation scenarios require a much higher space density of objects than is observed today. Additionally, most TNBs have identically-colored components, with differences between primary and secondary color being much smaller than the large color variation seen across the entire population of TNOs (Benecchi et al. 2009). This suggests that each component formed from similar material in a similar region of a locally homogeneous protoplanetary disk with global variations in composition. Different mechanisms proposed for binary formation dominate under different dynamical conditions (eg., Schlichting \& Sari 2008a). If the dynamical properties of the systems today can be taken to be representative of their primordial distribution, they can probe the dynamical conditions of the primordial Kuiper Belt during the formation phase. However, any intervening violent dynamical events, such as collisions (Petit \& Mousis 2004, Nesvorn\'{y} et al. 2011) or close encounters with giant planets (Parker \& Kavelaars 2010) can leave today's mutual orbit distribution substantially altered from its original state. It is critical to measure the orbital properties of a large sample of TNBs, as well as perform dynamical studies of possible sources of orbital modification, in order to understand the full extent of information about the formation and history of the outer Solar System encoded in these systems.

We have collected astrometric measurements of a sample of seven of the widest-known TNBs for an extended period, covering four to nine years of orbital motion for each system. These observations have allowed us to compute accurate mutual orbits for our sample of ultra-wide TNBs, and from these orbits we derive system mass and a host of other characteristics. In the first part of this paper, we outline the nomenclature we adopt to describe these systems and their host populations (\S 1.1), our sample selection criteria (\S 2), details of our observational campaign and data reduction techniques (\S 3), and mutual orbit fitting algorithm (\S 4). In the latter part, we describe the mutual orbit fits (\S 5) and compare them to the properties of other binary populations, and derive geometric albedos for each system given reasonable assumptions of bulk density (\S 6). Finally, we conclude with a discussion of possible formation mechanisms and implications for the early history of the outer solar system, susceptibility of these systems to disruption by collisions and Neptune scattering, and present future surveys' abilities to discover and characterize a large sample of these ultra-wide TNBs. 
\vspace{1 cm}

\subsection{Nomenclature}

In this paper, we compare several sub-populations of Trans-Neptunian Objects and their various orbital properties. In order to facilitate a clear understanding of the nomenclature we use to describe these populations and their properties, we provide an outline here.

A binary's mutual orbit properties will be described either as a ``mutual'' property or denoted by the subscript ``$m$.'' In contrast, the properties of the orbit of the binary's barycenter around the Sun will be described as an ``outer'' property or denoted by the subscript ``out.''

In order to compare the properties of our sample with those in the literature, some dynamical classification is required. We adhere roughly to the Gladman et al. (2008) nomenclature when discussing outer orbit properties. In this paper, we frequently deal with binaries that belong to the following dynamical classes:

\begin{itemize}

\item ``Classical:'' Non-resonant objects in the range 34 AU$\leq q_{\mbox{out}} \leq 47$ AU, 37 AU $\leq a_{\mbox{out}} \lesssim 70$ AU.

\item ``Cold Classical:'' Subset of ``Classical'' objects with low orbital excitations and confined in semi-major axis. When dividing samples, we assign ``Classical'' binaries with $i_{\mbox{out}} < 10^\circ$, $q_{\mbox{out}} > 38$ AU, and 42.4 $\leq a_{\mbox{out}} \leq 47$ AU to this population. Referred to as CC population in the text.

\item ``Hot Classical:'' Subset of ``Classical'' objects with higher mean orbital excitations, and an extension to lower pericenter than the CC population. When dividing samples, we assign ``Classical'' binaries with $i_{\mbox{out}} > 10^\circ$, $q_{\mbox{out}} < 38$ AU, $a_{\mbox{out}} < 42.4$ or $a_{\mbox{out}} > 47$ AU to this population. Referred to as HC population in the text.

\end{itemize}

This dynamical classification is somewhat different from that adopted by Grundy et al. (2011), and several binaries in that work which were classified as ``extended scattered'' fall into our HC classification.

In addition, we compare the outer orbital distributions of binary sub-samples with the Canada-France Ecliptic Plane Survey (CFEPS) L7 synthetic model of the Kuiper Belt\footnote{Available at \url{http://www.cfeps.net/L7Release.html}}. We compare the CC binary sub-sample with the composite of the ``stirred'' and ``kernel'' sub-components of the synthetic Kuiper Belt model, and refer to the composite of these sub-components as CC-L7. We compare the HC binary sub-sample with the ``hot'' sub-component of the synthetic Kuiper Belt model, and refer to this sub-component as HC-L7.

In reality, any simple inclination cut is insufficient to determine which population a given object truly belongs to, as both the CC and HC populations overlap significantly. According to the CFEPS L7 model, most relatively bright objects below $10^\circ$ of inclination actually belong to the HC-L7 population. We stress that while we will refer to a given object as a ``CC'' binary or an ``HC'' binary, there is no way to absolutely verify the parent population for a given single object. However, we show later that the sub-samples of binaries which fall into our CC and HC classifications have dynamically distinct mutual orbit distributions, and the outer orbit distributions of CC binaries suggest that they are in fact members of the CC-L7 population, and likewise the HC binaries' outer orbit distribution is consistent with the HC-L7 distribution.

\begin{table*}
\centering
\begin{centering}
\begin{tabular}{lccccccccc}
\multicolumn{10}{c}{\bf Table 1}\\
\multicolumn{10}{c}{Observations and System Properties}\\
\hline
Name & Date Range & $N_{visits}$ & $N_{obs}$ & $m_{sys}$ & $\Delta m$ & $H_{p}$$^{e}$ & \multicolumn{3}{c}{Outer Orbit}\\
            &                        &                &                             & ($r'$)                          &        ($r'$)     &   ($r'$) & $a_{\mbox{out}}$ (AU) & $e_{\mbox{out}}$& $i_{\mbox{out}}$ ($^{\circ}$)\\
\hline
\hline
2000 CF$_{105}$  & 2002-2011 &  12 & 50  & 23.85$^{a}$ & 0.72(5)   & 7.70 & 	43.84  	& 0.0362     	& 0.528 \\
2001 QW$_{322}$ & 2001-2010 &  35 & 88   & 23.16$^{b}$& 0.03(5)& 7.51 & 		43.98   	& 0.0242     	& 4.808 \\
2003 UN$_{284}$ &  2003-2010 & 14 & 60  & 22.7$^{c}$ & 0.88(6) & 7.5 &  		42.62   	& 0.0035     	& 3.069\\
2005 EO$_{304}$ & 2005-2011 &  12  & 52  & 22.45$^{d}$ & 1.45(3)& 6.59 & 		45.62   	& 0.0679     	& 3.415 \\
2006 BR$_{284}$  & 2006-2011 &                       20  & 66  & 23.0$^{a}$  & 0.50(4)& 7.3 & 				43.80   	& 0.0393     	& 1.157 \\
2006 JZ$_{81}$    & 2006-2011 &                      15 & 56  & 22.7$^{a}$  & 0.98(2)& 6.9 &  				44.70   	& 0.0804     	& 3.550\\
2006 CH$_{69}$   & 2004-2010 &                      15 & 47  & 23.0$^{a}$  & 0.44(5)& 7.0 & 				45.74   	& 0.0362     	& 1.791 \\
\hline
\end{tabular}
\end{centering}
\flushleft
{\footnotesize \hspace{2cm} {\bf Notes.}\\
\hspace{2cm} $^{a}$: From CFHT MegaPrime, using \textit{Elixir} photometric solutions. \\
\hspace{2cm} $^{b}$: From Gemini North, Petit et al. (2008). \\
\hspace{2cm} $^{c}$: From Gemini North, this work. \\
\hspace{2cm} $^{d}$: From KPNO Mayall telescope, Benecchi et al. (2009). \\
\hspace{2cm} $^{e}$: Assuming phase correction of 0.14 magnitudes per degree.}

\end{table*}

\section{Sample Selection}

Since we seek to characterize the widest binaries (which have correspondingly long mutual periods), we opted to pursue a ground-based observation campaign. We chose our sample based on the following criteria: 

\begin{enumerate}
\item The system had no well-characterized orbit in literature.
\item The separation at discovery was $\gtrsim0\arcsec.5$.
\item The magnitude difference between the system's primary and secondary was less than 1.7, indicating a near-equal mass system ($M_p / M_s \lesssim 10$). 
\end{enumerate}

At the time of our sample selection, there were seven systems that met these criteria: 2000 CF$_{105}$, 2001 QW$_{322}$, 2003 UN$_{284}$, 2005 EO$_{304}$,  and three objects discovered over the course of CFEPS, with internal designations b7Qa4, L5c02, and hEaV. The binary nature of 2000 CF$_{105}$ was presented in Noll et al. (2002), while the binary natures of 2003 UN$_{284}$ and 2005 EO$_{304}$ were presented in Kern (2006). Provisional orbital characterization for 2001 QW$_{322}$ was presented in Petit et al. (2008). The CFEPS target L5c02 was identified as binary by Lin et al. (2010), and the binary nature of b7Qa4 and hEaV are presented here for the first time. All of the outer orbits of this sample of objects fall into our CC classification, and have very low outer inclinations and eccentricities. 

Two other CFEPS targets, L4q10 and L4k12, were initially included in our sample, due to data from the Canada-France-Hawaii Telescope (CFHT) suggesting that they were elongated in a manner consistent with a near-equal mass binary with a separation of order $0\arcsec.5$. However, follow-up observations in very good seeing did not bear out their putative binary nature, and they were removed from our target list. The three CFEPS objects in our sample also have MPC designations: b7Qa4 is 2006 BR$_{284}$, hEaV is 2006 JZ$_{81}$, and L5c02 is 2006 CH$_{69}$. Throughout this paper we will refer to these three systems via their MPC titles.

\section{Observations and Data Reduction}

A targeted observational campaign from 2008---2011 was executed from \textit{Gemini North} using the \textit{Gemini Multi-Object Spectrograph} in imaging mode, taken with the $r_{G0303}$ filter. Observations were queue scheduled, with stringent requirements on image quality (frequently at the expense of photometric conditions). By requiring modest visit times ($\sim30$ minutes), excellent seeing could be obtained without the use of adaptive optics, in some cases with full-width half-max of $\Gamma \simeq 0\arcsec.35$ or better. Additional observations during this period were made from \textit{VLT} with the \textit{FORS2} instrument, though image quality requirements were not held to the same stringent limits. Single-epoch observations were also made in April 2010 from \textit{Magellan} with the \textit{Megacam} imager.

Significant archival data also exists for all systems. We used the \textit{Solar System Object Search} (S.D.J. Gwyn 2011, in prep) service provided by the \textit{Canadian Astronomy Data Centre}\footnote{\url{http://www2.cadc-ccda.hia-iha.nrc-cnrc.gc.ca/ssos/}} to locate and download images from the \textit{CFHT} and \textit{HST} public archives that contained our targets, and we also located images of our targets from the \textit{Mayall}, \textit{Hale}, and \textit{WIYN} telescopes. 

Literature astrometric measurements are available for some systems as well. The relative astrometry for 2001 QW$_{322}$ published in Petit et al. (2008) is also included in our fit for that system. Astrometric measurements of 2003 UN$_{284}$ and 2005 EO$_{304}$ were presented in Kern (2006), and we include those measurements in our fits for these systems.

Combining all these data sources, the smallest number of observed epochs for any binary in our sample is 12 visits for 2005 EO$_{304}$, while the largest number is 35 visits for 2001 QW$_{322}$. During most visits, more than one usable image was acquired. The number of visits and total number of images from which astrometric measurements were made are listed in Table 1. 

Astrometric solutions were generated for each individual image, matched to the J2000 coordinate system using reference stars in the \textit{USNO b}  astrometric catalog. Whenever possible, the catalog stellar positions were corrected for proper motion since their last observed epoch, and uncertainties in the final reference positions reflected the original astrometric precision and the integrated uncertainty due to the stated uncertainty in proper motion over the intervening time.

The brightest 100 non-saturated stars were identified in the CCD on which the binary was located (in the case of multi-chip imagers, other chips in the array were not used to constrain the astrometric solution), and their ($x,y$) positions (and uncertainties) extracted using \textit{SExtractor} (Bertin \& Arnouts, 1996). In the case of an image with a good initial World Coordinate System (WCS) estimate (eg., \textit{Gemini GMOS} images), this WCS was used to estimate each stars' RA and DEC position in the J2000 system and the nearest neighbor in the \textit{USNO b} catalog was identified as its matching counterpart.

If an image did not have a good initial WCS, a robust pattern-recognition algorithm identified probable rough corrections to the WCS, applied these corrections, and then identified nearest-neighbor stars in the \textit{USNO b}  catalog.

Once \textit{USNO b}  (RA, DEC) positions were matched to ($x,y$) positions in the image, the \textit{IRAF} package \textit{ccmap} was used to generate a WCS solution. Because this package does not handle positional uncertainties in either the image or reference positions, the positions of each matched star is cloned 1,000 times, adding Gaussian noise to each position consistent with the (RA, DEC) and ($x,y$) uncertainties. Iterative fitting followed by automatic clipping of outliers in these thousands of cloned sources allowed \textit{ccmap} to generate a robust astrometric solution automatically which reflected the uncertainties in the absolute and measured positions of the reference stars. This allowed a more robust automatic solution to be derived with little input from the operator for each image processed. 

After the first pass of \textit{ccmap}, more matches are searched for in the image with the \textit{USNO b} catalog reference stars, and upon flagging any new matches the \textit{ccmap} routine is called again. This matching and WCS-fitting process is iterated ten times for every frame.

\begin{figure}
\begin{centering}
\includegraphics[width=0.5\textwidth]{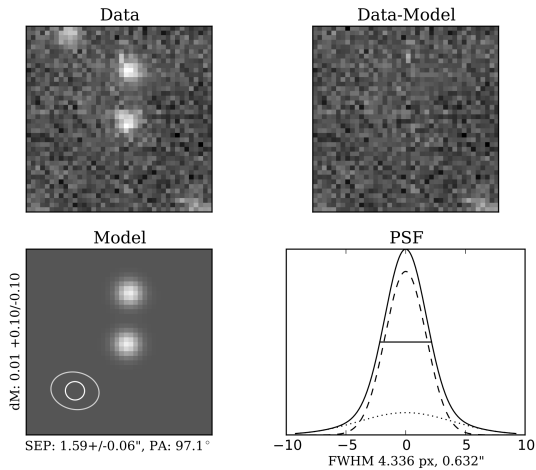}
\caption{Example of Gemini data and PSF fit. Top left: Original image from GMOS camera of 2001 QW$_{322}$. Bottom left: Synthetic PSF model of binary components. Top right: Image residuals after subtracting binary and other point-sources in the image. Bottom right: Relative contributions of both PSF components. Same stretch is applied to all images, and flux scaling is linear.}
\label{example_data}
\end{centering}
\end{figure}

\begin{figure*}
\begin{centering}
\includegraphics[width=0.45\textwidth]{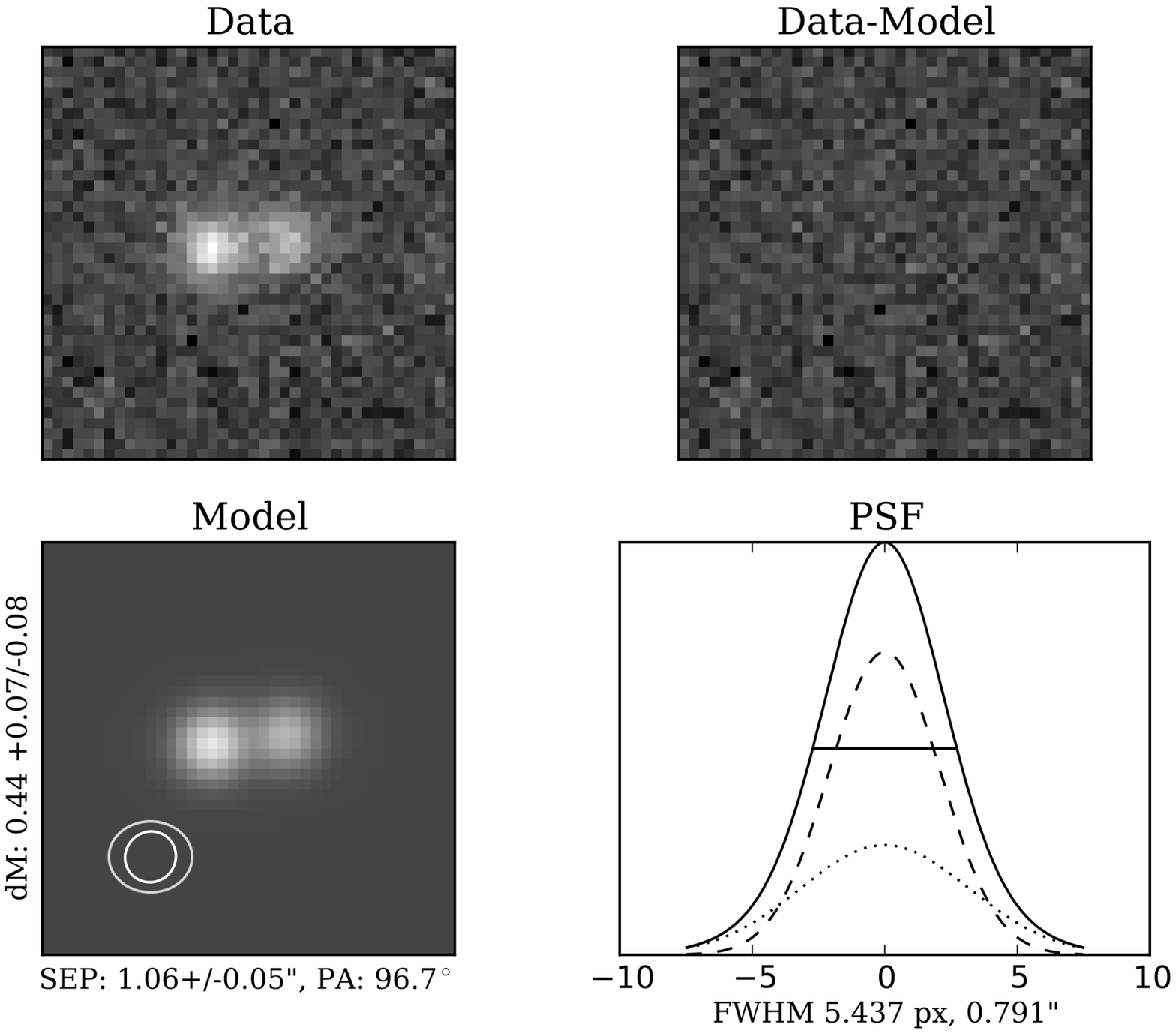}
\includegraphics[width=0.45\textwidth]{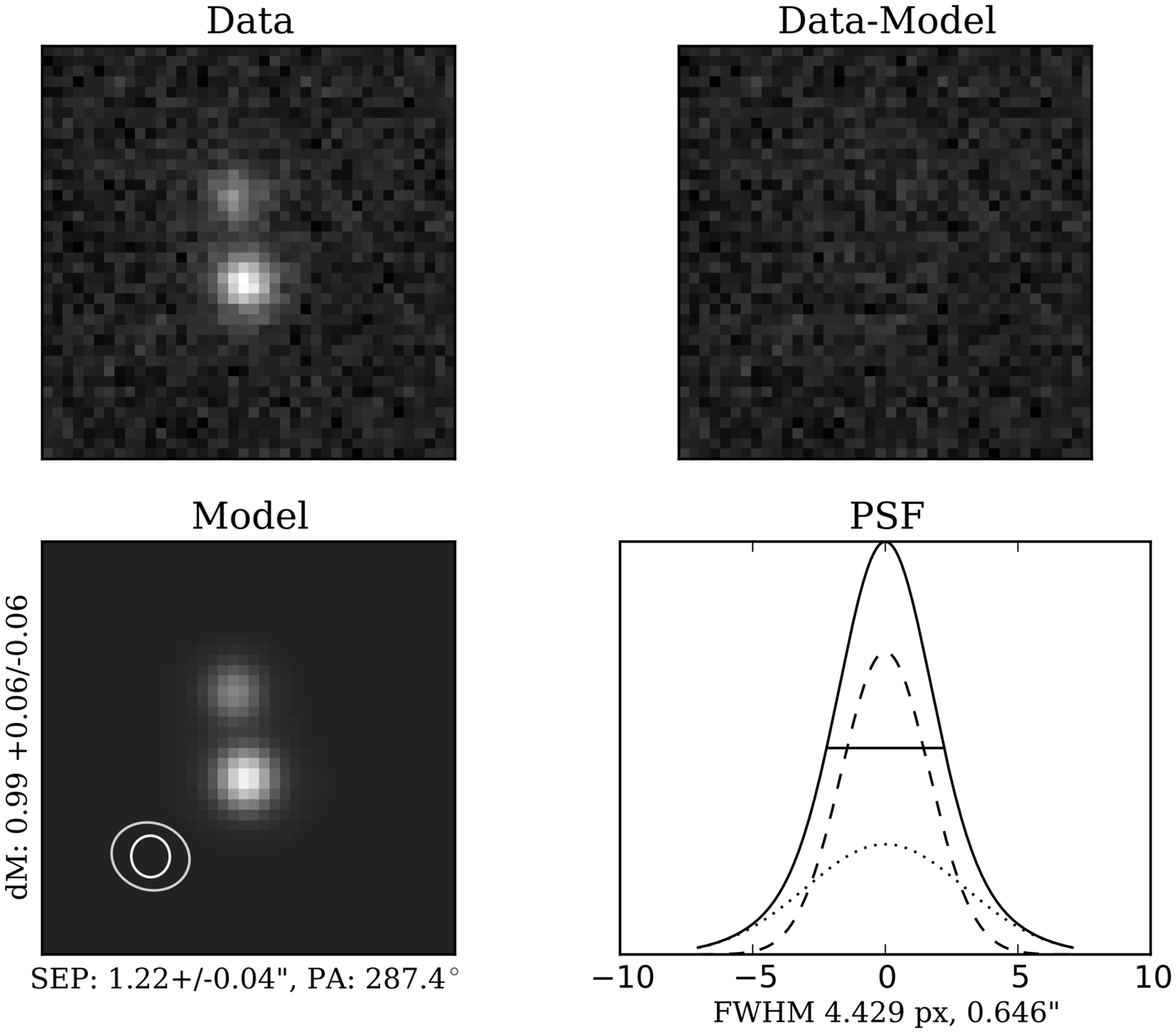}
\caption{Same as Figure \ref{example_data}, but for CFEPS binaries 2006 BR$_{284}$ (left) and 2006 JZ$_{81}$ (right).}
\label{example_data2}
\end{centering}
\end{figure*}

The lowest-order astrometric solution merited by the distortions of the optics of each imager was used in each case. In the case of \textit{Gemini GMOS} images, this was a simple rotation and fixed pixel scale. In the case of most other imagers, the distortion across the area of a single CCD was low enough such that the addition of independent $x$ and $y$ pixel scales, as well as a skew term, was sufficient. In the case of HST observations, the internal HST astrometric solutions and distortion corrections were used.

Once an astrometric solution had been found for an image, the relative astrometry of the binary pair in that image was then extracted using a custom Point-Spread Function (PSF) fitting routine. The PSF model we adopted was a sum of two elliptical Gaussian components, with arbitrary long-axis orientation for each component. The wider of the two Gaussians (the PSF wings) is arbitrarily limited to contain less than $2/3$ the flux of the narrower ``core'' Gaussian to prevent runaway solutions with extremely wide wings. Adopting a non-circular PSF was required due to the fact that most images were obtained with sidereal tracking rates, and over the period of integration the PSF of the binary's components became somewhat stretched along their direction of motion. 

A variant of the same algorithm used for fitting the mutual binary orbits (described in the following section) was used to minimize the residuals in a sky-subtracted 40$\times$40 pixel region centered on the binary. An initial interactive step is used to identify all the point sources in this region and flag the two associated with the binary. Initial estimates for the amplitude and $\Gamma$ of the PSF are made automatically, and these values are fed into a Markov-Chain Monte Carlo algorithm which finds the PSF model and array of point source positions and amplitudes that produces the smallest residuals. Because the image model varies with the number of point sources in the 40$\times$40 pixel region, the minimum number of free parameters the algorithm must search over is 9 (two point sources, and a two-component PSF model forced to be circular) while the maximum number of free parameters ever treated was 22 (five point sources and a two-component PSF model with arbitrary rotation and ellipticity for each component). Example of data from the \textit{Gemini} observatory is illustrated in Figures 1 and 2, along with the image residuals after PSF fitting and subtraction, and the model of the binary system. Each fit is visually inspected, and in general we found that our adopted PSF model produced extremely low residuals.

In some cases where the two binary components were blended, we performed a check to verify that the extra degrees of freedom added by allowing the PSF to be elongated was not skewing the measured astrometry. In these cases, we performed a second, independent fit using a circular PSF and compared the measured astrometry for the binary components. In general, we found excellent agreement between the two. In cases of very strong blending, fits with the elongated PSF would occasionally have trouble converging to a stable solution, and in these cases we adopted the values from the circular PSF fits. We did not attempt to determine any upper limits on separation based on completely unresolved images, and therefore at present such images do not contribute to our mutual orbit fits.

Relative astrometry was recorded as separation (in arcseconds) and position angle (in degrees East of North), and the observation date was taken at the central Julian Date of each observation. Uncertainty in the relative astrometry was estimated as $\sigma_{xy} \simeq \Gamma \sqrt{ \mbox{SN}_p^{-2} + \mbox{SN}_s^{-2} }$, where $SN$ indicates signal-to-noise ratio of primary or secondary. A noise floor is set by the uncertainty in the astrometric solution for each frame. 

These PSF fits also returned relative photometry for each system, and the mean $\Delta$-mag measured in the \textit{Gemini GMOS} $r_{G0303}$ filter during well-resolved visits is also included in Table 1. Relatively few observations were made in photometric conditions, as image quality was our primary concern. All MPC targets in our sample have published absolute system photometry in various bands, but only 2001 QW$_{322}$ and 2005 EO$_{304}$ have $r'$-band magnitudes in literature. 2000 CF$_{105}$, 2006 BR$_{284}$, 2006 JZ$_{81}$, and 2006 CH$_{69}$ were all imaged on photometric nights from CFHT, and \textit{Elixir} processed images were used to determine $r'$-band system magnitudes for these systems. The $r'$-band magnitude of 2003 UN$_{284}$ was determined from observations on a single night from Gemini North, though the absolute calibration of these particular images is poor and the resulting photometric uncertainty is relatively large. 

%\begin{equation}
%\chi^2 = \sum\limits_{x,y}^{}  \frac{ \left( \, I\!(x,y) - M\!(x,y) \, \right)^2 }{ | \, I_0\!(x,y) \, |}
%\end{equation}

\section{ Mutual Orbit Determination }

The basic operations performed by our mutual-orbit fitting routine are, given an initial guess of mutual orbit properties, to solve Kepler's equation in order to determine the relative system geometry at the time of observation (accounting for the light-time delay between the system and observer), then rotate the system in space to account for its orientation with respect to the Ecliptic. Finally, the code ``observes'' the system by applying a second rotation to account for the variation in viewing geometry induced by the relative motion between the Earth and the binary, and projects the result onto the sky plane given the separation between the observer and the system. 

To fit our observations, we chose to adopt the Metropolis algorithm $\chi^2$ minimization routine (Metropolis et al. 1953), using a similar implementation to that described in Simard et al. (2002), who utilized the Metropolis algorithm to fit a 12 dimensional bulge + disk model to images of galaxies. This algorithm is robust to complicated topology in parameter space, and can be easily adjusted to thoroughly explore parameter space at the expense of speed. The Metropolis algorithm is a Markov-Chain Monte Carlo technique which, after an initial burn-in period, occasionally makes ``bad'' decisions, allowing it to diffuse out of local minima. After a number of iterations, the choice of new parameter values can be informed by previous values, improving the speed of convergence in complicated parameter-space topology. A binary mutual orbit has seven free parameters, and in our implementation we chose these to be the following: mutual semi-major axis ($a_m$), eccentricity ($e_m$), period ($T_m$), Mean Anomaly ($M$, valid at a defined JD), inclination ($i_E$), longitude of the ascending node ($\Omega_E$), and argument of pericenter ($\omega_E$). The last three angular parameters are defined with respect to the J2000 Ecliptic. For nearly circular orbits, $M$ and $\omega$ become degenerate and alternate choice of basis is preferred; however, we were not presented with a circumstance where altering the basis used in our code was merited, as the binaries we observed all have significant eccentricity. For all orbit fits, 15 Metropolis algorithm threads are run simultaneously, and each compare their final best-fit value and sampled orbit space to identify the global best-fit and statistically acceptable range of parameters after an additional test of the error distribution.

\begin{table}
\centering
\begin{centering}
\begin{tabular}{ccccc}
\multicolumn{5}{c}{\bf Table 2}\\
\multicolumn{5}{c}{Mutual Astrometry}\\
\hline
Name & JD & Sep & PA & Err \\
     & (d) & ($\arcsec$) & ($^\circ$) & (\arcsec) \\
\hline
\hline
2001QW322&2452117.92296&3.6056&290.0270&0.2030 \\
2001QW322&2452145.74085&3.8915&291.3190&0.1722 \\
2001QW322&2452145.76181&3.8683&287.1030&0.1845 \\
2001QW322&2452145.78289&3.9600&290.0990&0.1168 \\
2001QW322&2452146.75689&3.7377&292.2420&0.2255 \\
\hline
\end{tabular}
\end{centering}

{\footnotesize
Table 2 is published in its entirety in the electronic \\
edition of the {\it Astrophysical Journal}.  A portion is \\
shown here for guidance regarding its form and content.}
\end{table}

Because our estimates of the astrometric uncertainty of each observation may not reflect their true distribution, (i.e., we don't know the properties of the distribution of errors on the individual measurements \textit{a priori}), we perform a test to bootstrap the 68\% and 95\% confidence intervals for the $\chi^2$ distribution. This test simply assumes that the best-fit orbit is the ``truth,'' and estimates the observed error distribution around the best-fit. The test randomly draws $n$ measurements of the observed error from the best-fit orbit from the pool of $n$ real observations of a given target (sampled with replacement, so observations may be repeated in the resampled list). After this resampling, we compute a new ``observed'' $chi^2$ based off of the resampled list and best-fit orbit. We store this new $chi^2$ statistic and repeat the process 10,000 times, building up a distribution of best-fit $chi^2$ values which ``might have been.'' We find the 68\% and 95\% upper confidence intervals on this distribution, and set those to be our $\chi^2$ thresholds for statistically acceptable mutual orbit fits. Orbits that fall below these $\chi^2$ thresholds are saved, and their distribution is used to generate the uncertainties for each orbital parameter. This analysis has led to extremely well-behaved fitting behavior, with consistently nested uncertainty contours after every addition of new data to a given fit.

%After each resampling, a new ``observed'' $\chi^2$ is calculated based off of the resampled observations and the best-fit orbit. After this process, we are left with a distribution of possible best-fit $\chi^2$ values which ``might have been.'' 

The mutual orbit fitting code was tested by reproducing the orbital parameters for the Pluto-Charon system based on synthetic data generated by the \textit{JPL Horizons} system, and reproducing the orbital parameters of 2001 XR$_{254}$ and 2004 PB$_{204}$ as published in Grundy et al. (2009) to an accuracy well within their stated uncertainties. We did not explore the robustness of our method for determining confidence intervals with these tests, as we did not have sufficient numbers of real observations for the two published mutual orbits we tested.

\section{ Present Best-Fit Orbits and Implications }

\begin{figure*}[t]
\begin{centering}
\includegraphics[width=0.45\textwidth]{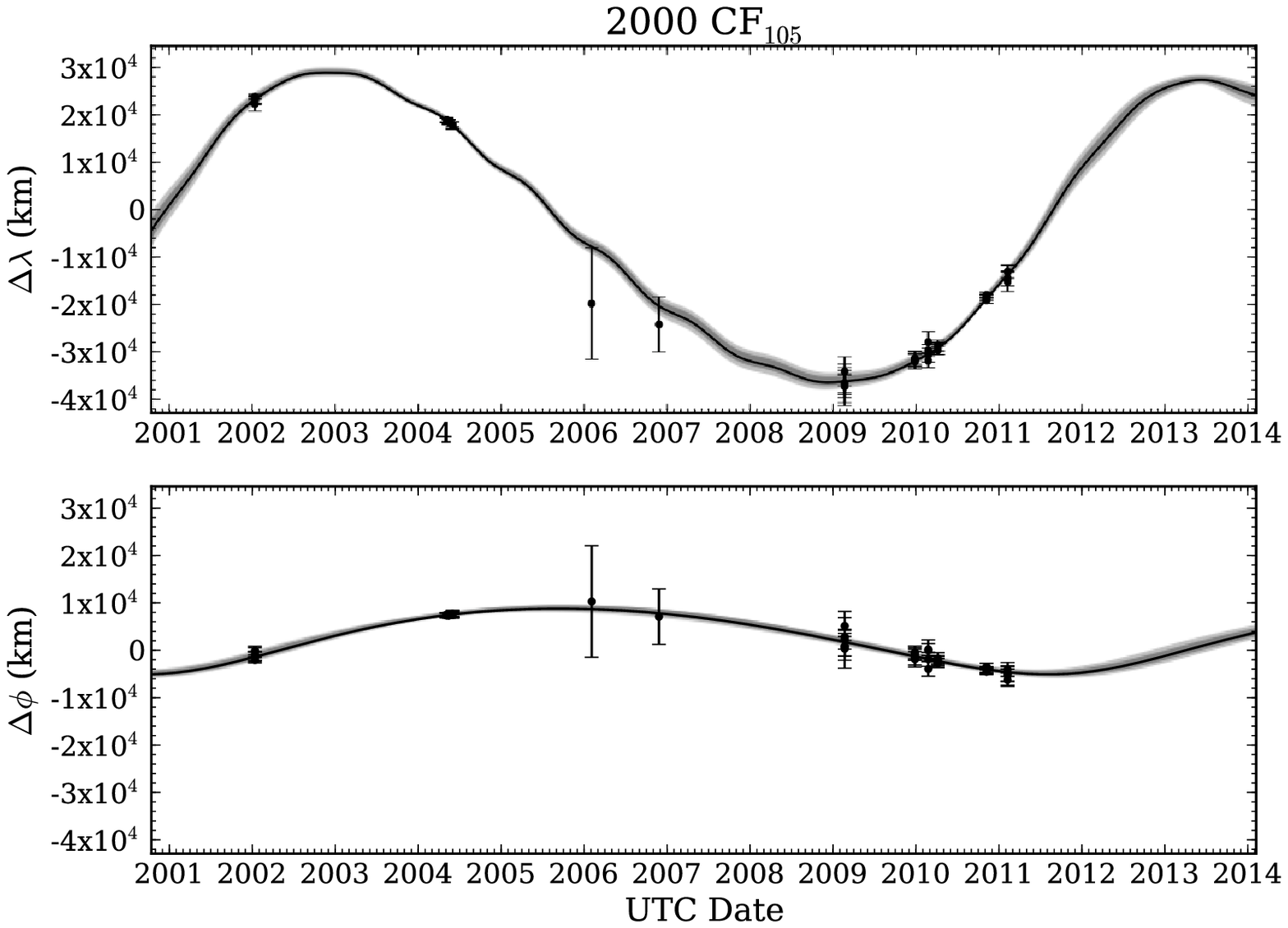}
\includegraphics[width=0.45\textwidth]{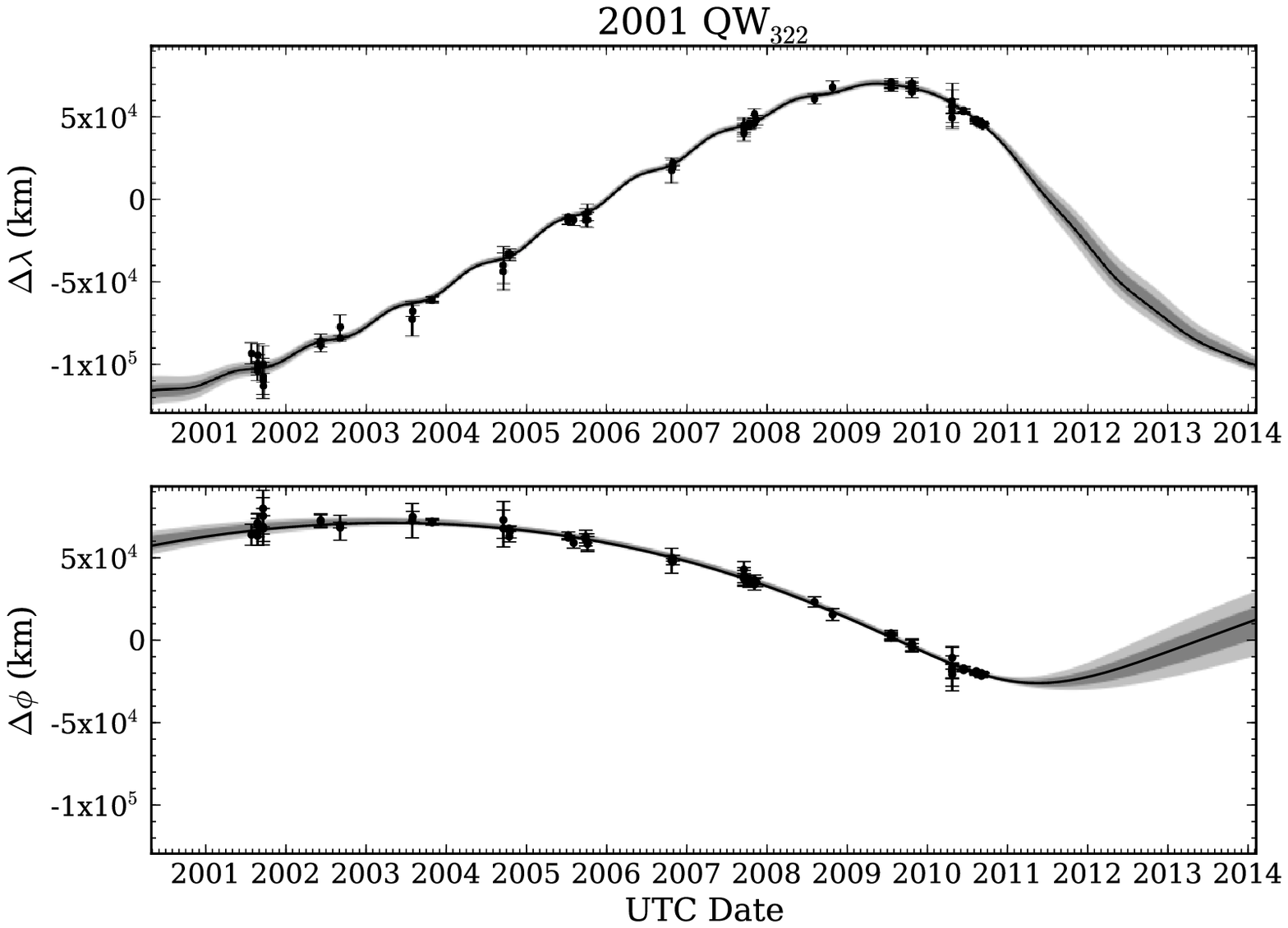}
\includegraphics[width=0.45\textwidth]{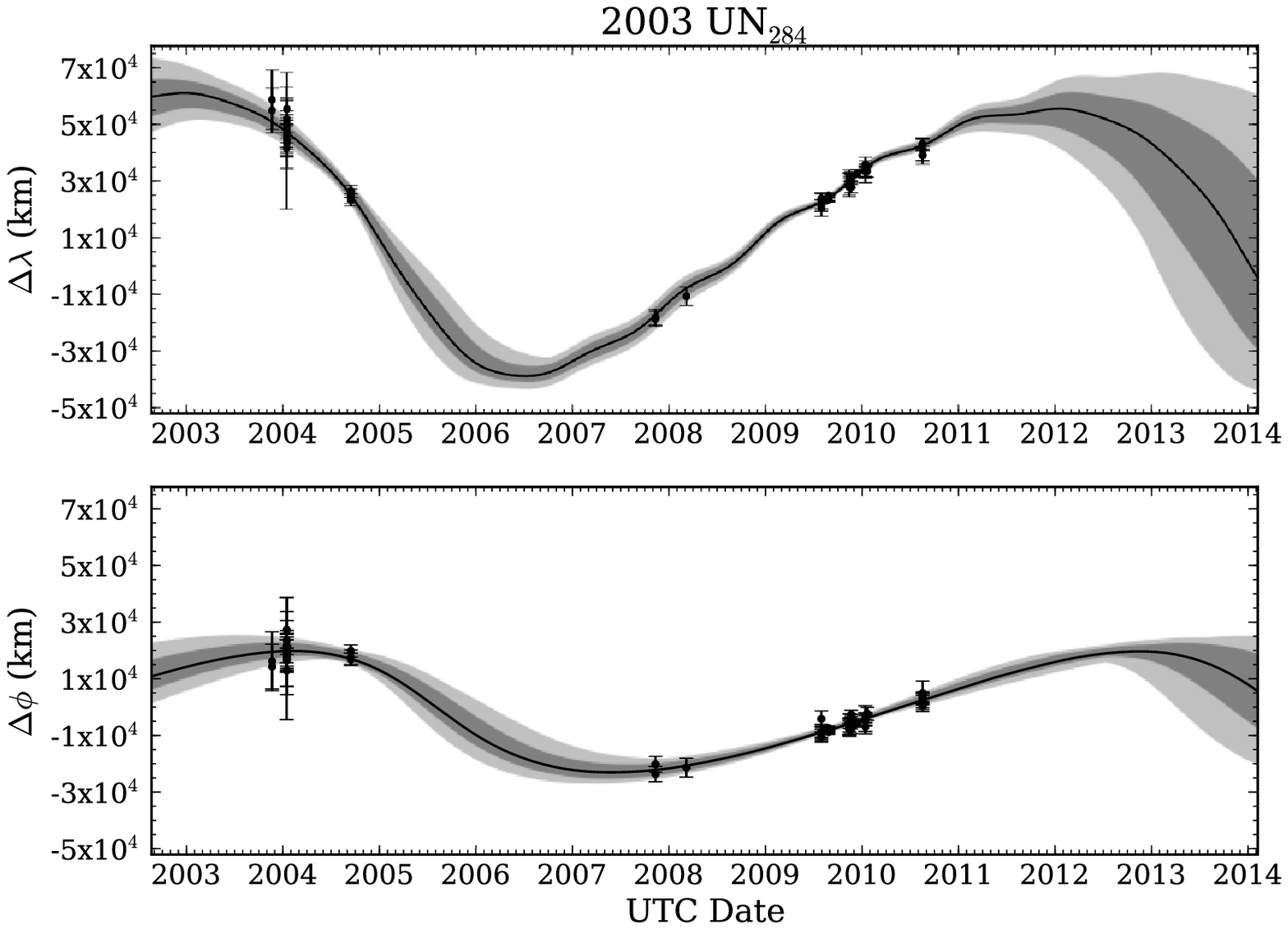}
\includegraphics[width=0.45\textwidth]{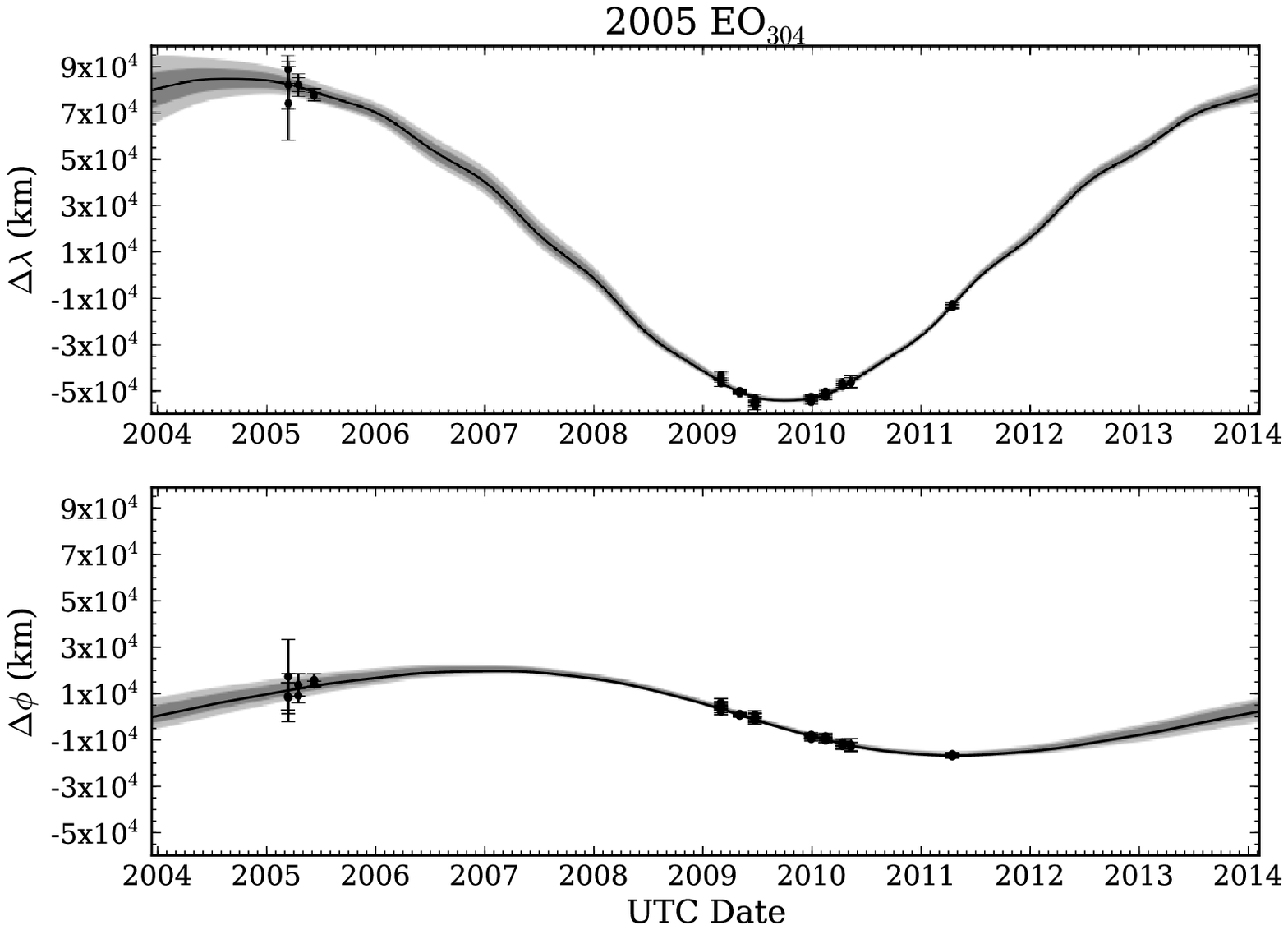}
\caption{Astrometry and fitted mutual orbits for MPC binaries 2000 CF$_{105}$, 2001 QW$_{322}$, 2003 UN$_{284}$, and 2005 EO$_{304}$. Latitude separation ($\Delta\phi$) and longitude separation ($\Delta\lambda$) are given in projected physical units (km) to remove variation due to changing separation between binary system and the observer and illustrate physical scale of each system. Black line indicates best-fit mutual orbit, while dark and light gray regions illustrate orbits consistent at the 68\% and 95\% confidence level, respectively. Mutual astrometry is available online as a machine-readable table.}
\label{MPC_binaries}
\end{centering}
\end{figure*}

\begin{figure*}
\begin{centering}
\includegraphics[width=0.45\textwidth]{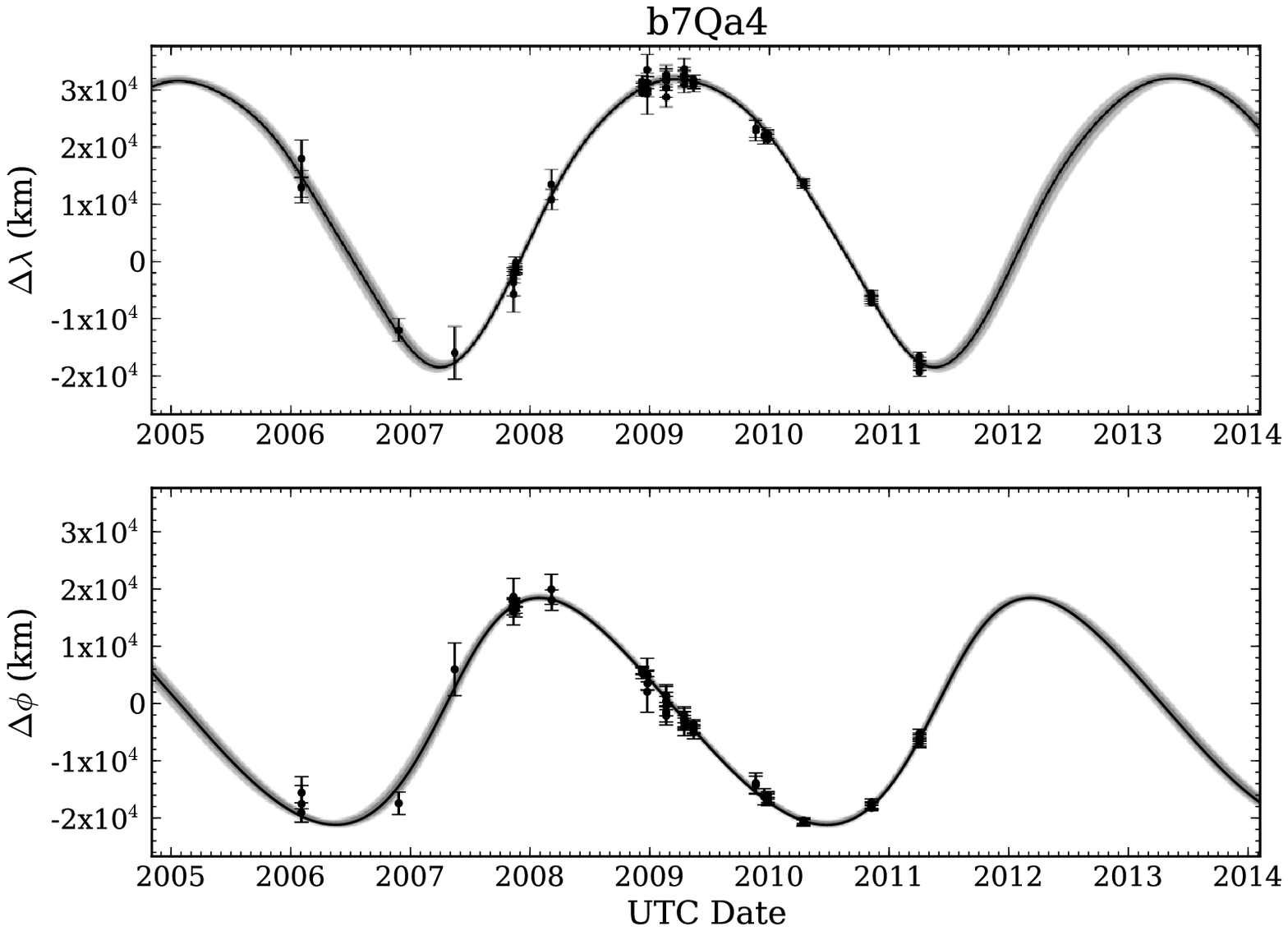}
\includegraphics[width=0.45\textwidth]{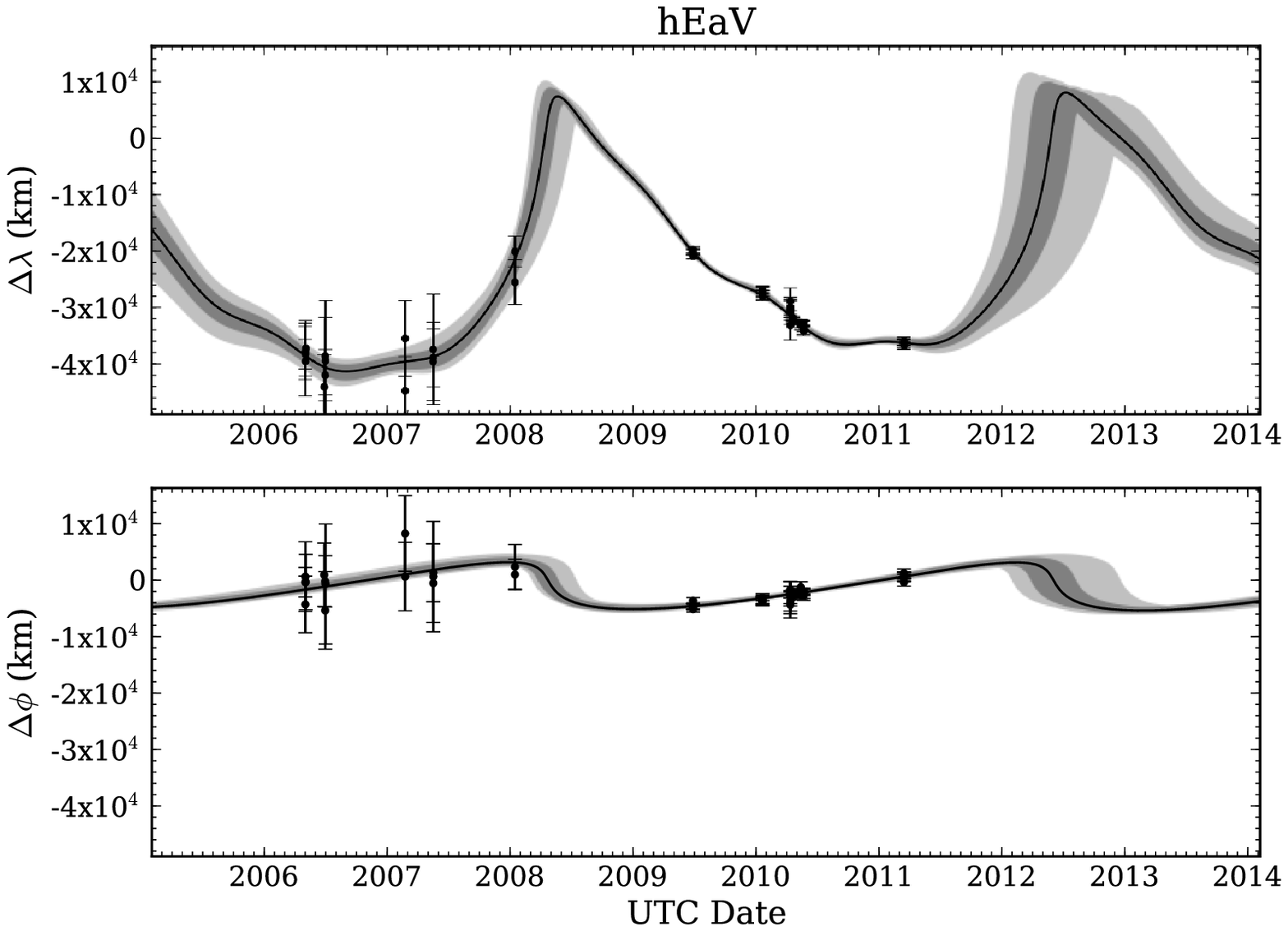}
\includegraphics[width=0.45\textwidth]{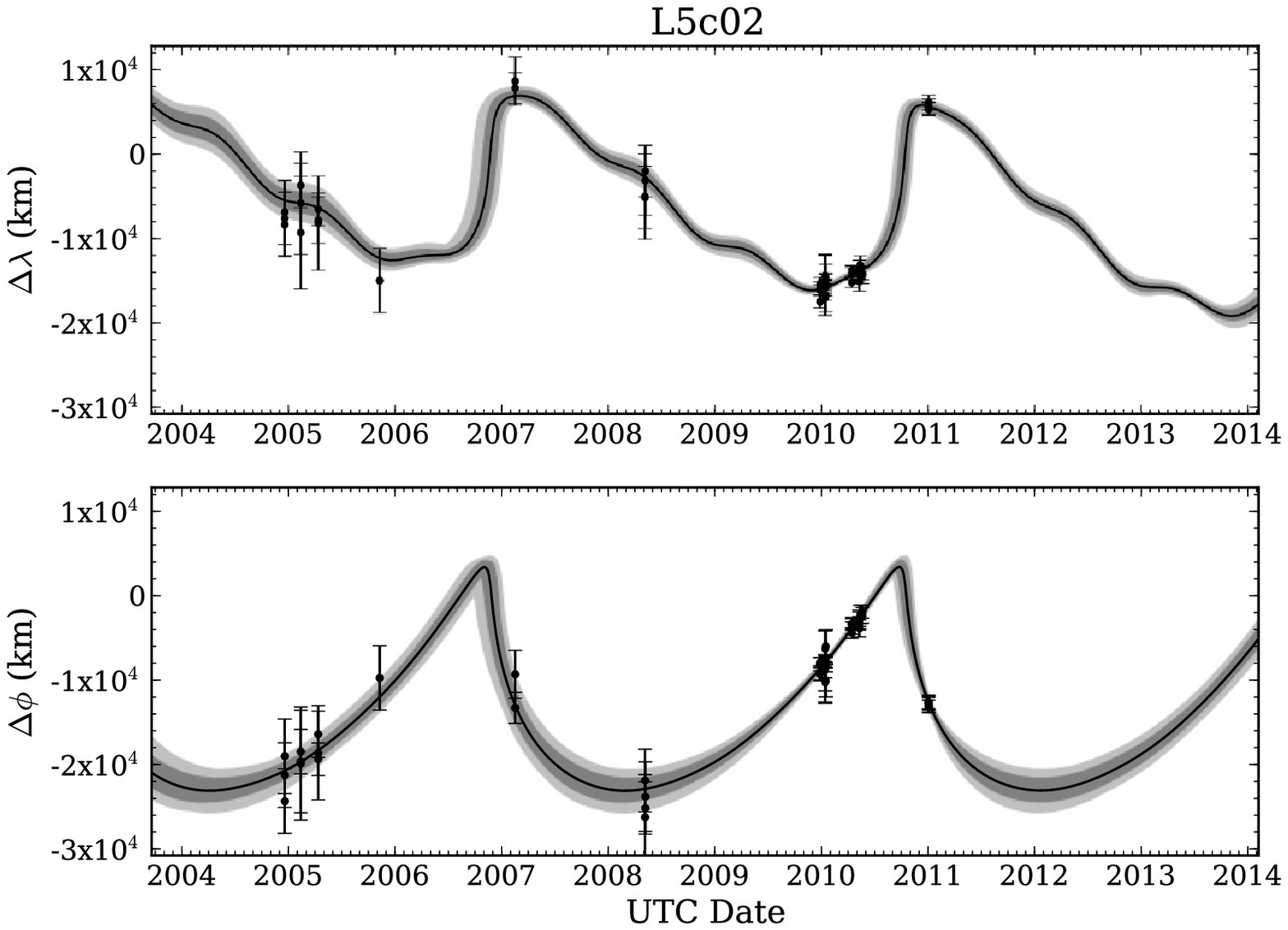}
\caption{Same as Figure \ref{MPC_binaries}, but for the three CFEPS binaries 2006 BR$_{284}$ (b7Qa4), 2006 JZ$_{81}$ (hEaV), and 2006 CH$_{69}$ (L5c02). Mutual astrometry is available online as a machine-readable table.}
\label{CFEPS_binaries}
\end{centering}
\end{figure*}

The astrometric measurements for each system and comparison to fit orbits are illustrated in Figures \ref{MPC_binaries} and \ref{CFEPS_binaries}. These figures project each system onto the sky plane in physical separation units, removing the variation in the observed scale of the systems due to the change in the observer-system separation over the course of a year.
%% Newly added Table 2 reference here %%
An example of the  mutual astrometry for all the TNBs used in these 
Figures is shown in Table 2.  The full table is available in the electronic
edition.

The best-fit and uncertainty (given as the extrema of each parameter from the distribution of orbits allowed at the 68\%  level of confidence) of all fitted mutual orbit parameters are listed in Table 3. Additionally, Table 4 contains derived parameters; specifically, the system mass $M_{sys}$, mutual semi-major axis to Hill-radius fraction $a/R_H$, mutual inclination $i_{m}$, mutual argument of pericenter $\omega_{m}$, and mutual pericenter separation in multiples of primary radii $q_m / R_p$. Figure \ref{polar_contour} illustrates the $a/R_H$, $e_m$, and $i_m$ fits and the 68\% and 95\% uncertainties in these parameters for each system.

\begin{table*}
\centering
\begin{tabular}{ l r@{}l r@{}l c r@{}l  r@{}l r@{}l r@{}l c }
\multicolumn{15}{c}{\bf  Table 3}\\
\multicolumn{15}{c}{Fit Mutual Orbit Elements}\\
\hline
Name & \multicolumn{2}{c}{$a_m$} & \multicolumn{2}{c}{$T_m$}   & $e_m$ & \multicolumn{2}{c}{$i_E$} & \multicolumn{2}{c}{$\Omega_E$} & \multicolumn{2}{c}{$\omega_E$} &  \multicolumn{2}{c}{$M$} &  Epoch\\
&  \multicolumn{2}{c}{($10^4$ km)} & \multicolumn{2}{c}{(years)}  & &  \multicolumn{2}{c}{($^\circ$)}  &  \multicolumn{2}{c}{($^\circ$)} &  \multicolumn{2}{c}{($^\circ$)} & \multicolumn{2}{c}{($^\circ$)} \\
\hline
\hline
2000 CF$_{105}$ 	& $3.$ 	& $33^{+0.05}_{-0.06}$ 		& $10.$	& $92^{+0.12}_{-0.10}$ 	& $0.29^{+0.02}_{-0.02}$ & $167.$ & $4^{+0.6}_{-0.7}$ 	& $223.$ & $^{+4}_{-3}$ 	& $296.$ & $^{+3}_{-3}$ 		& $262.$ & $^{+3}_{-3}$ 		& 2454880.96\\
2001 QW$_{322}$ 	& $10.$ 	& $15^{+0.38}_{-0.14}$ 		& $17.$	& $01^{+1.55}_{-0.69}$ 	& $0.46^{+0.02}_{-0.01}$ & $150.$ & $7^{+0.6}_{-0.6}$ 	& $243.$ & $^{+3}_{-4}$ 	& $257.$ & $^{+5}_{-10}$ 	& $158.$ & $^{+19}_{-10}$ 	& 2452117.92\\
2003 UN$_{284}$ 	& $5.$ 	& $55^{+0.38}_{-0.53}$ 		& $8.$	& $73^{+0.65}_{-0.54}$ 	& $0.40^{+0.04}_{-0.07}$ & $24.$ & $3^{+2.2}_{-1.5}$ 	& $92.$ & $^{+6}_{-3}$ 	& $172.$ & $^{+10}_{-8}$ 	& $294.$ & $^{+5}_{-14}$ 	& 2452963.77\\
2005 EO$_{304}$ 	& $6.$ 	& $98^{+0.20}_{-0.21}$ 		& $9.$	& $80^{+0.45}_{-0.45}$ 	& $0.22^{+0.02}_{-0.02}$ & $12.$ & $4^{+0.8}_{-0.5}$ 	& $259.$ & $^{+2}_{-3}$ 	& $206.$ & $^{+9}_{-5}$	 	& $193.$ & $^{+8}_{-13}$ 	& 2453440.94\\
2006 BR$_{284}$      			& $2.$ 	& $53^{+0.03}_{-0.03}$ 		& $4.$	& $11^{+0.04}_{-0.03}$ 	& $0.28^{+0.01}_{-0.01}$ & $55.$ & $6^{+1.3}_{-1.4}$ 	& $41.$ & $^{+2}_{-2}$ 	& $14.$ & $^{+1}_{-1}$ 		& $219.$ & $^{+1}_{-1}$ 		& 2455153.08\\
2006 JZ$_{81}$ 			& $3.$ 	& $23^{+0.53}_{-0.28}$ 		& $4.$	& $11^{+0.15}_{-0.12}$ 	& $0.84^{+0.03}_{-0.02}$ & $13.$ & $3^{+2.5}_{-1.9}$ 	& $82.$ & $^{+5}_{-7}$ 	& $171.$ & $^{+2}_{-2}$ 		& $104.$ & $^{+9}_{-8}$ 		& 2455007.86\\
2006 CH$_{69}$ 			& $2.$ 	& $76^{+0.33}_{-0.28}$ 		& $3.$	& $89^{+0.05}_{-0.07}$ 	& $0.90^{+0.02}_{-0.02}$ & $134.$ & $1^{+4.9}_{-6.1}$	& $105.$ & $^{+6}_{-8}$ 	& $149.$ & $^{+5}_{-6}$ 		& $286.$ & $^{+4}_{-3}$ 		& 2455190.06\\
\hline
\end{tabular}
\end{table*}

\begin{table*}
\centering
\begin{tabular}{l  r@{}l r@{}l r@{}l r@{}l r@{}l}
\multicolumn{11}{c}{\bf Table 4}\\
\multicolumn{11}{c}{Derived Values}\\
\hline
Name &  \multicolumn{2}{c}{$M_{sys}$} &  \multicolumn{2}{c}{$a_m / R_H$} &  \multicolumn{2}{c}{$i_m$}  &  \multicolumn{2}{c}{$\omega_m$} &  \multicolumn{2}{c}{$q_m/R_P$$^*$} \\
&  \multicolumn{2}{c}{($10^{17}$ kg)} & & &  \multicolumn{2}{c}{($^\circ$)} &  \multicolumn{2}{c}{($^\circ$)} & & \\
\hline
\hline
2000 CF$_{105}$	& $1.$ & $85^{+0.1}_{-0.14}$ 		& $0.$ & $1679^{+0.0012}_{-0.0011}$ 	& $167.$ & $9^{+0.6}_{-0.7}$  	& $295.$ & $^{+3}_{-3}$ 			& $741$ & $^{+29}_{-30}$\\
2001 QW$_{322}$	& $21.$ & $50^{+1.44}_{-2.23}$ 	& $0.$ & $2222^{+0.0133}_{-0.0061}$ 	& $152.$ & $7^{+0.6}_{-0.8}$  	& $248.$ & $^{+6}_{-10}$ 		& $855$ & $^{+64}_{-40}$\\
2003 UN$_{284}$ 	& $13.$ & $12^{+2.26}_{-2.97}$ 	& $0.$ & $1449^{+0.0070}_{-0.0060}$ 	& $22.$   & $7^{+2.2}_{-1.4}$ 	& $165.$ & $^{+21}_{-8}$ 		& $534$ & $^{+59}_{-42}$\\
2005 EO$_{304}$ 	& $21.$ & $03^{+0.87}_{-0.74}$ 	& $0.$ & $1553^{+0.0048}_{-0.0047}$ 	& $15.$ & $7^{+0.8}_{-0.5}$  	& $203.$ & $^{+9}_{-5}$	 		& $714$ & $^{+7}_{-6}$\\
2006 BR$_{284}$ 			& $5.$ & $70^{+0.17}_{-0.20}$ 		& $0.$ & $0879^{+0.0005}_{-0.0005}$ 	& $54.$ & $6^{+1.3}_{-1.4}$  	& $13.$ & $^{+1}_{-2}$ 			& $408$ & $^{+8}_{-7}$\\
2006 JZ$_{81}$ 			& $11.$ & $83^{+7.09}_{-3.18}$ 	& $0.$ & $0900^{+0.0021}_{-0.0018}$ 	& $11.$ & $1^{+2.5}_{-2.0}$  	& $158.$ & $^{+4}_{-4}$ 			& $84$ & $^{+10}_{-18}$\\
2006 CH$_{69}$ 			& $8.$ & $30^{+3.35}_{-2.15}$ 		& $0.$ & $0809^{+0.0007}_{-0.0009}$ 	& $133.$ & $3^{+4.9}_{-4.8}$  	& $147.$ & $^{+5}_{-6}$ 			& $56$ & $^{+10}_{-10}$\\
\hline
\end{tabular}

{\footnotesize$^*$:  Primary radius $R_P$ assumes $\rho=1$ gram cm$^{-3}$.}
\end{table*}

\subsection{Derived Parameters}

\begin{figure*}
\begin{centering}
\includegraphics[width=\textwidth]{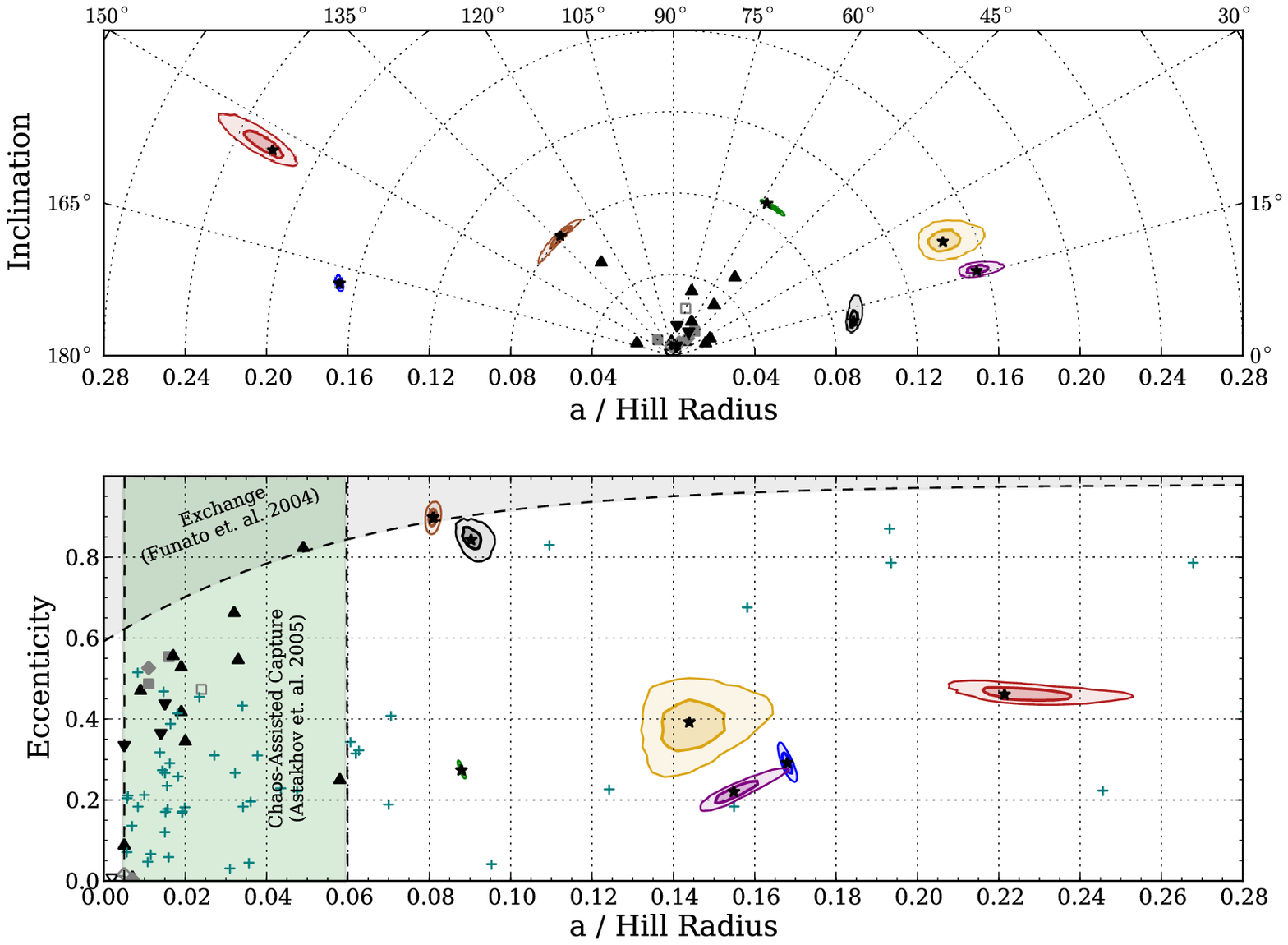}
\caption{Best-fit mutual orbit properties. Blue: 2000 CF$_{105}$, Red: 2001 QW$_{322}$, Yellow: 2003 UN$_{284}$, Purple: 2005 EO$_{304}$, Green: 2006 BR$_{284}$, Gray: 2006 JZ$_{81}$, Brown: 2006 CH$_{69}$. Heavy and light contours represent 68\% and 95\% confidence intervals, respectively, while black stars mark the best-fit parameters. Teal ``+'' symbols mark sample of synthetic systems created by gravitational collapse in Nesvorn\'{y} et al. (2010), selecting simulations with $\Omega=0.5-1.0\Omega_{circ}$ (initial clump rotation), and $f^*=10-30$ (cross-section modifier). Gray shaded region represents properties of binary systems created by three-body exchange reactions in Funato et al. (2004). Green shaded region represents properties of binary systems created by Chaos-Assisted Capture in Astakhov et al. (2005). Other points represent TNBs with known orbits (Grundy et al. 2011). Black points are members of the Classical belt, gray points are members of other dynamical populations. Filled points indicate $\Delta m < 1.7$, and open points indicate $\Delta m > 1.7$. Upward triangles indicate CC objects, while downward triangles indicate HC objects. Square points indicate resonant objects, and diamond points indicate Centaurs or scattered disk objects. Non-preferred degenerate pole solutions are not illustrated for literature binaries.}
\label{polar_contour}
\end{centering}
\end{figure*}

In the following section, we describe the derived mutual orbital parameters listed in Table 4 and illustrated in Figure \ref{polar_contour}. System mass is simply calculated from Kepler's laws, 

\begin{equation}
M_{sys} = 4\pi^2 \frac{ a_{m}^3 }{ G T_m^2 }.
\end{equation}

\noindent while the classical Hill radius for a binary in orbit around the Sun is defined as 

\begin{equation}
R_H = a_{\mbox{out}} (1-e_{\mbox{out}})\left( \frac{M_{sys}}{3 M_{\odot}} \right)^{\frac{1}{3}},
\end{equation}

\noindent where $a_{\mbox{out}}$ and $e_{\mbox{out}}$ is are the heliocentric semi-major axis and eccentricity of the binary system's barycenter. Primary radius is found from the system mass by assuming that both components have the same albedo and density, and is given by

\begin{equation}
R_p = \left( \frac{3}{4\pi\rho( 1 + 10^{\frac{-3\Delta m}{5}})} M_{sys} \right)^{\frac{1}{3}},
\end{equation}

\noindent where $\rho$ is the adopted bulk density and $\Delta m$ is the magnitude difference between the primary and secondary components of the binary.

Mutual inclination is the angle between the pole vector of the binary mutual orbit $\vec{P_m}$ and that of the outer orbit $\vec{P}_{\mbox{out}}$, and can be found by $i_m = \cos^{-1}(\vec{P}_{\mbox{out}} \cdot \vec{P_m})$. The pole vectors for either orbit can be found by (eg., Naoz et al. 2010):

\[  \vec{P}= \left( \begin{array}{cc}
&\sin(\Omega)\sin(i)\\
-\!&\!\cos(\Omega)\sin(i)\\
&\cos(i) \end{array} \right).\]

The mutual argument of pericenter $\omega_{m}$ (critical for estimating the extent of Kozai oscillations) is the angle between the ascending node (with respect to the outer orbit) and pericenter in the plane of the mutual orbit. It can be found by:

\begin{equation*}
\omega_{m} = \cos^{-1} \left( \frac{ \vec{e}_m \cdot \vec{n} }{ |\vec{e}_m| |\vec{n}|}\right)\mbox{sign}(\vec{e}_m[z] ),
\end{equation*}

\noindent where

\[  \frac{ \vec{e}_m}{ | \vec{e}_m | } = \left( \begin{array}{c}
\cos(\omega_E)\cos(\Omega_E)\!-\!\sin(\omega_E)\cos(i_E)\sin(\Omega_E) \\
\cos(\omega_E)\sin(\Omega_E) \!+\! \sin(\omega_E)\cos(i_E)\cos(\Omega_E) \\
\sin(\omega_E)\sin(i_E) \end{array} \right),\] 

\noindent and $\vec{n} = \vec{P}_{\mbox{out}} \times \vec{P_m}$.

%%%%%%%%%%%%%%%%%%%%%

\subsection{Kozai Cycles}

Systems within a broad range of $i_m$ centered on $90^\circ$ may be subject to large oscillations in $i_m$ and $e_m$ due to the Kozai effect (Kozai 1962). Over the oscillation cycles two values are conserved: one depending on the initial mutual eccentricity and inclination, and the other depending on both these and the mutual argument of pericenter. Following Perets \& Naoz (2008), we adopt the following form for these two conserved values:

\begin{equation}
A = \left( 5e_m^2\sin^2\omega_m + 2(1-e_m^2) \right)\sin^2i_m
\end{equation}

\noindent and

\begin{equation}
B = \sqrt{ 1 - e_m^2 } \cos i_m.
\end{equation}

With some algebraic manipulation and the constraint that eccentricity and inclination minima and maxima occur when $\omega_m = 0^\circ$ or $90^\circ$, these two constants determine the maximum and minimum eccentricity and inclination reached by a given binary system during its Kozai cycle. We calculate these values, and the amplitudes of the eccentricity excursions experienced by each binary system is listed in Table 5. Also listed are the predicted minimum pericenter separations (occurring during the highest eccentricity phase of the Kozai cycle) in multiples of primary radii, as close encounters during Kozai cycles may lead to modification of the mutual orbit due to tidal friction (Fabrycky \& Tremaine 2007, Perets \& Naoz 2009, Brown et al. 2010). Details of systems with large-amplitude Kozai oscillations will be discussed on an object-by-object basis in the following sections.

The Kozai effect can be easily suppressed by other effects, including permanent asymmetries in the mass distribution of the component bodies (Ragozzine 2009). However, this suppression only occurs for relatively small semi-major axes, and therefore the Kozai oscillations of these wide systems are unlikely to be suppressed.

\subsection{Individual Objects}

\subsubsection{2000 CF$_{105}$}

With several epochs of HST data, and nearly nine years of observational baseline (covering most of a single 11 year mutual orbit period), 2000 CF$_{105}$ has one of the best-measured orbits in our sample with mutual semi-major axis and period uncertainties at the 1\% level. It is also the lowest-mass system in our sample, and the second-most weakly bound (behind only 2001 QW$_{322}$). Because of its low mass (currently the lowest mass of any known TNB), this system has the smallest estimated primary radius of any system in our sample (assuming all objects share a common density), estimated to be $31.8^{+0.6}_{-0.8}$ km given a bulk density of 1 gram cm$^{-3}$. 

2000 CF$_{105}$ is not subject to strong Kozai oscillations, and its pericenter separation is always greater than $\sim710$ primary radii, so there is little chance of mutual tides having modified its mutual orbit. 

The pole solution for 2000 CF$_{105}$ is non-degenerate at greater than 95\% confidence, and the system is retrograde. Its mutual pole vector is only $\sim12.1^\circ$ degrees from anti-aligned with its outer orbit's pole vector, making it one of the most pole-parallel systems known.

\subsubsection{2001 QW$_{322}$}

\begin{figure}
\begin{centering}
\includegraphics[width=0.5\textwidth]{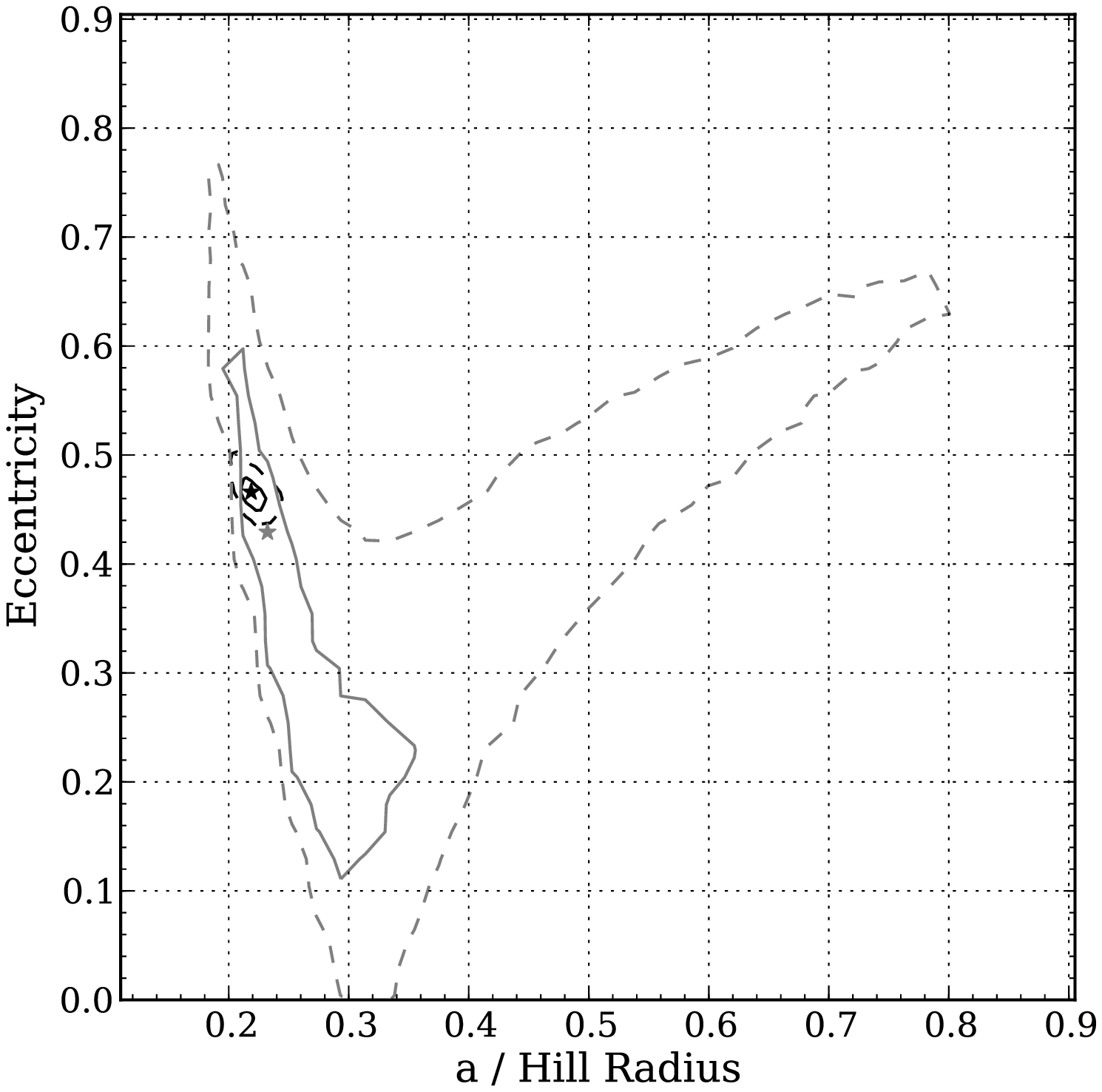}
\caption{Comparison of the allowed orbits for 2001 QW$_{322}$, given the data presented in Petit et al. 2008 (gray contours) and the complete set of data used in this work (black contours) using the fitting algorithm described in the text. Stars mark best-fit orbits. 68\% and 95\% confidence intervals are marked by the solid and dashed contours, respectively.}
\label{comp_QW322}
\end{centering}
\end{figure}

At the outset, our current results for the mutual orbit of 2001 QW$_{322}$ appear inconsistent with the orbit published in Petit et al. (2008). By fitting only the astrometric data used in that paper with our new code, we find a much larger range of allowable mutual orbit solutions than was presented in the previous study. This is especially notable for allowing much higher values of mutual eccentricity. This larger range of allowed orbital parameters completely overlaps with our current orbit fit, as illustrated in Figure \ref{comp_QW322}. We suspect that the difference is due to two factors: our more thorough fitting algorithm and our different statistical analysis of the allowed $\chi^2$ range. The allowed $\chi^2$ range in Petit et al. (2008) was much smaller than that determined by the methods used in this work. 

The components of this system remain photometrically indistinguishable, with $\Delta m$ consistent with 0. When making astrometric measurements, we have arbitrarily assigned the Northernmost object in the discovery epoch as the system primary. Observations have been frequent enough that there is no possibility for confusion between system components, based on continuity of the orbital motion.

The observations in our dataset cover approximately nine years, sampling slightly over 50\% of the best-fit mutual orbital period of 17.0 years. Because the observations have been frequent and of high quality, this limited sample of the orbital motion of the system is very constraining. Mutual semi-major axis, period, and eccentricity are all known to better than 10\% accuracy. Angular and derived parameters are similarly well known.

2001 QW$_{322}$ is subject to strong Kozai oscillations, with the nominal best-fit orbit implying eccentricity variations between $0.342 \lesssim e_m \lesssim 0.477$. Its current orbit is therefore near its highest eccentricity phase.

In physical units, this system remains the most widely separated binary minor planet known, with a mutual semi-major axis of $1.015^{+0.038}_{-0.014} \times 10^{5}$ km. It is also likely the most weakly bound binary minor planet known, with its measured $a/R_H$ of $0.2222^{+0.0133}_{-0.0061}$ exceeding the current estimate for the outer satellite of the Main-Belt asteroid (3749) Balam of $a/R_H \sim 0.2$ (Marchis et al. 2008). Several other known main-belt asteroid binaries have estimates of $a/R_H$ shown in  Richardson \& Walsh (2006) which are similar to or slightly higher than that measured for 2001 QW$_{322}$, but these estimates are based on single-epoch observations and may not reflect the true orbits and masses of these systems.

The pole solution for 2001 QW$_{322}$ is non-degenerate at greater than 95\% confidence, and the system is retrograde with a mutual inclination of $\sim152.7^\circ$.

\subsubsection{2003 UN$_{284}$}

The orbit of 2003 UN$_{284}$ is the least-well constrained in our sample. The fit relies heavily on astrometry published in Kern (2006) for pinning the 2003---2004 astrometry, and only three data points constrain the 2005---2008 astrometry. Recent data has proved relatively discriminatory, and the mutual semi-major axis and period are both known to better than 10\% accuracy, but the derived system mass is only constrained to $\sim20$\% accuracy. The $\Delta m$ we adopt for 2003 UN$_{284}$ is determined from two well-resolved visits from Gemini North. Kern (2006) finds a highly variable $\Delta m$ for this system, suggesting one or both components may have significant lightcurve, and our adopted $\Delta m$ may not reflect this variability.

This system is likely subject to minor Kozai oscillations, but the amplitude of these oscillations are comparable to the uncertainty in the currently measured mutual eccentricity. The pole solution for 2003 UN$_{284}$ is non-degenerate at greater than 95\% confidence, and the system is prograde with a mutual inclination of $\sim22.7^\circ$.

\subsubsection{2005 EO$_{304}$}

The orbit of 2005 EO$_{304}$ also relies heavily on astrometry published in Kern (2006), and it has the fewest observed epochs (12) of any of our systems. Nevertheless, the recent measurements from \textit{VLT} and \textit{Gemini} have provided reasonably tight constraints on the orbit properties.

This system has the largest $\Delta m$ in our sample at 1.45 in $r'$. It also has the largest primary radius (assuming all objects share a common density), which we estimate to be $76.2^{+1.0}_{-0.9}$ km given a bulk density of 1 gram cm$^{-3}$.

2005 EO$_{304}$ is not subject to significant Kozai oscillations, and its minimum pericenter separation is at least 692 primary radii. The pole solution for 2005 EO$_{304}$ is non-degenerate at greater than 95\% confidence, and the system is prograde with a mutual inclination of $\sim15.7^\circ$, making it one of the lowest mutual inclination systems known.

\subsubsection{2006 BR$_{284}$}

Nearly all observations of 2006 BR$_{284}$ used in our astrometric fit come from \textit{Gemini}, and the orbit has been sampled for just over a single orbital period. Its orbit is very well constrained, with mutual semi-major axis and period known to 1\% accuracy and mutual eccentricity to better than 4\% accuracy. The pole solution for 2006 BR$_{284}$ is non-degenerate at greater than 95\% confidence, and the system is prograde with a mutual inclination of $\sim54.6^\circ$, making it the most inclined system in our sample.

2006 BR$_{284}$ is subject to strong Kozai oscillations, with the nominal best-fit orbit implying eccentricity variations between $0.72 \leq e_m \leq 0.26$. Its current orbit is therefore near the lowest point of its eccentricity cycle. Its pericenter passages are still widely separated (never lower than 143 primary radii) and mutual tides are not a concern for this system.

\subsubsection{2006 CH$_{69}$}

\begin{figure}
\begin{centering}
\includegraphics[width=0.5\textwidth]{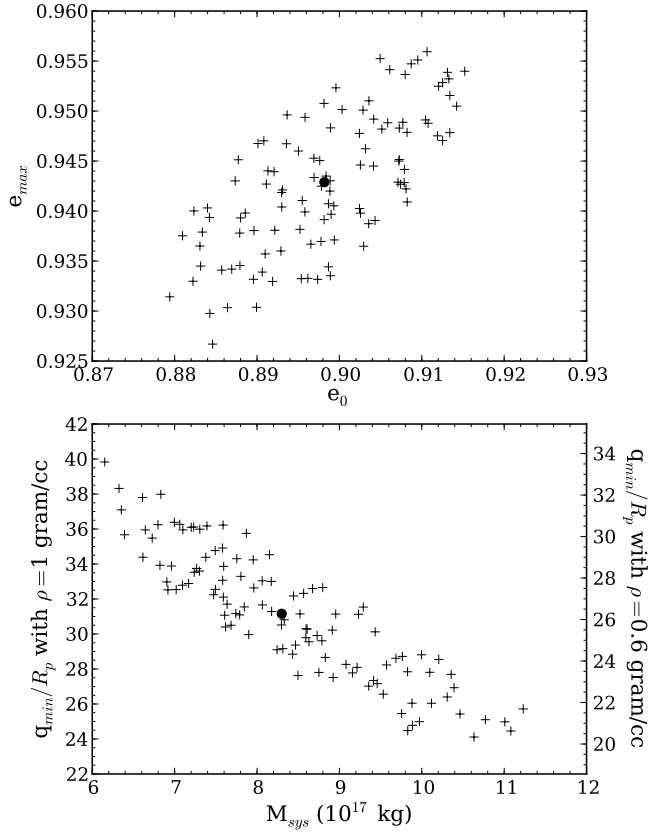}
\caption{Features of 2006 CH$_{69}$'s Kozai oscillations. Top panel: representative sample of 107 orbits consistent with 2006 CH$_{69}$ astrometry at $1\sigma$ level, showing their current eccentricity ($e_0$) and the maximum eccentricity ($e_{max}$) they reach over the course of a Kozai cycle, with the best-fit orbit marked by the large point. Bottom panel: Same sample of orbits, but now illustrating their best-fit system mass ($M_{sys}$) and the minimum pericenter separation in multiples of primary radii ($q_{min}/R_p$, assuming $\rho=1$ gram cm$^{-3}$).}
\label{L5c02_kozai}
\end{centering}
\end{figure}

The observations of 2006 CH$_{69}$ cover well over a single orbital period, though as Figure \ref{CFEPS_binaries} illustrates its on-sky behavior is quite complex and well time-sampled observations were necessary to accurately constrain this system's mutual orbital properties. 

This system has the highest well-measured mutual eccentricity of any binary minor planet, at $e_m = 0.90^{+0.02}_{-0.02}$. The orbit of the outermost satellite of the trinary asteroid (3749) Balam has been estimated to rival this at $e_m \sim 0.9$ (Marchis et al. 2008), but its value is poorly constrained. A fascinating consequence of this extreme mutual eccentricity is that over the course of its mutual orbit, the secondary of 2006 CH$_{69}$ will subtend an angle ranging from to $\sim0.1^\circ$ (at mutual apocenter) to $\sim1.7^\circ$ (at mutual pericenter) as viewed from the surface of the primary --- from one-fifth to over three times the angular size of the Moon on the sky.

2006 CH$_{69}$ may be subject to strong Kozai oscillations, with the nominal best-fit orbit implying eccentricity variations between $0.70 \lesssim e_m \lesssim 0.94$. Its current orbit is therefore near its highest eccentricity phase. The pole solution for 2006 CH$_{69}$ is non-degenerate at greater than 95\% confidence, and the system is retrograde with a mutual inclination of $\sim134^\circ$. This system is also the most tightly bound in our sample, with $a/R_H \simeq 0.0809$. Because of the orbits' high eccentricity, the error distributions in $a_m$ and $T_m$ conspire to produce a large uncertainty in the derived system mass, and $M_{sys}$ remains uncertain at the 40\% level.

The high mutual eccentricity phases of 2006 CH$_{69}$'s Kozai cycles lead to very close passages between the primary and secondary. At its current mutual eccentricity, pericenter passages occur at $56^{+10}_{-10}$ primary radii (again given a bulk density of 1 gram cm$^{-3}$), which may be wide enough that mutual tides do not cause orbital modification. However, during the high eccentricity phase of its Kozai cycle, the pericenter separation of 2006 CH$_{69}$ drops much lower to $31^{+9}_{-7}$. Figure \ref{L5c02_kozai} illustrates the distribution of mutual pericenter separation versus system mass. If we argue that 2006 CH$_{69}$ must have survived roughly in its current orbital configuration for the age of the solar system, then we would prefer higher minimum pericenter separations to keep the system from experiencing tidal evolution. From Figure \ref{L5c02_kozai} we see that orbit fits which have higher pericenter separations are lower-mass solutions. However, the tidal evolution of highly eccentric, highly inclined binary systems is poorly understood, and future work is merited to determine if limiting the tidal evolution of 2006 CH$_{69}$ would provide useful priors for further constraining its mutual orbit.

\subsubsection{2006 JZ$_{81}$}

The observations of 2006 JZ$_{81}$ span slightly more than a single period, and Figure \ref{CFEPS_binaries} illustrates that like 2006 CH$_{69}$ the on-sky behavior of this system is also complex. This system is also very highly eccentric, at $e_m = 0.84^{+0.03}_{-0.02}$, making it the second-highest eccentricity TNB known behind 2006 CH$_{69}$. Its mutual semi-major axis remains somewhat poorly constrained at 16\% uncertainty, while the mutual period is known to better than 5\% uncertainty. The derived system mass remains highly uncertain for the same reason as 2006 CH$_{69}$, and uncertainties in $M_{sys}$ remain at the 40\% level.

The pole solution of 2006 JZ$_{81}$ is non-degenerate at the 95\% level, and the system is prograde with the lowest mutual inclination of any known TNB at just $\sim11^\circ$. Due to this low inclination, 2006 JZ$_{81}$ is not subject to notable Kozai oscillations. Despite its high eccentricity and relatively large primary ($61^{+10}_{-6}$ km), mutual pericenter passages are always at least 64 primary radii, making this system much less susceptible to possible mutual tidal effects than 2006 CH$_{69}$.

\subsection{Ensemble Results}

\subsubsection{Comparison to Other Populations}

\begin{figure*}[t]
\begin{centering}
\includegraphics[width=\textwidth]{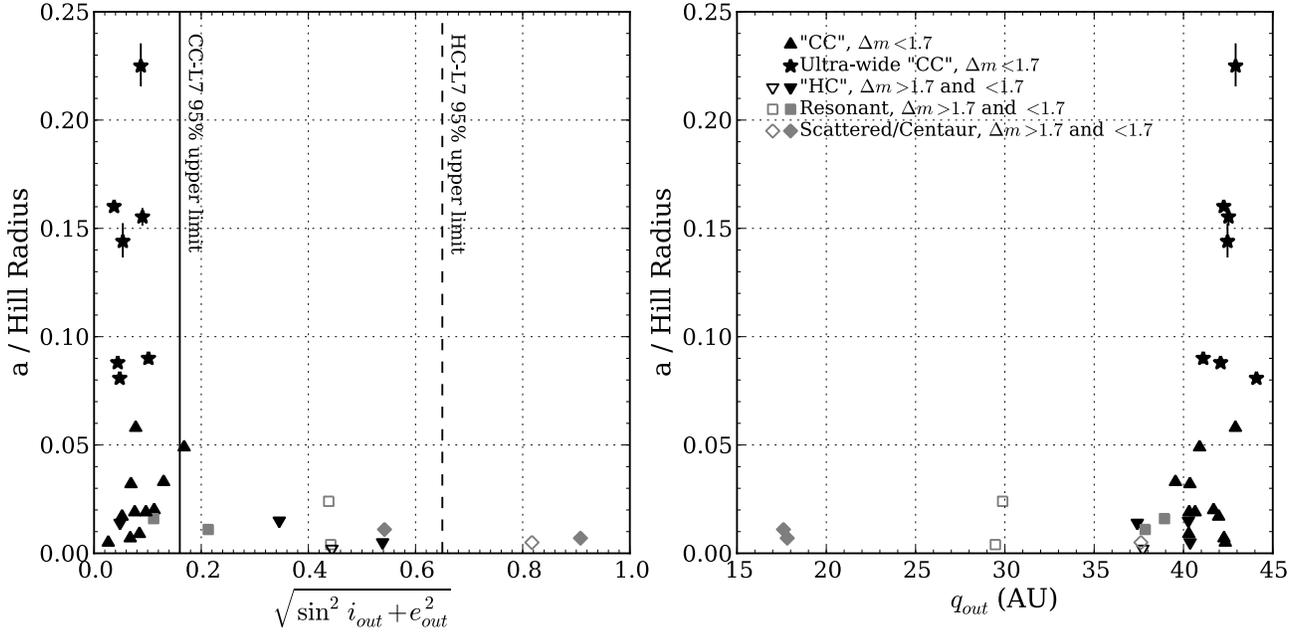} %%% fig 8
\caption{Left panel: Heliocentric orbital excitation vs. $a/R_H$ (similar to Fig. 5 from Grundy et al. 2011). Triangle, square, and diamond points have same meaning as in Figure \ref{polar_contour}, representing orbits published in Grundy et al. (2011). Stars mark best-fit orbits for the ultra-wide TNBs characterized in this work, with errorbars representing 68\% confidence interval. Vertical lines mark upper 95th-percentile of the orbital excitation of the cold Classical Kuiper Belt (solid) and hot Classical Kuiper Belt (dashed), as found by the CFEPS L7 survey. Right panel: heliocentric pericenter $q_{out}$ vs. $a/R_H$. }
\label{helio_excite}
\end{centering}
\end{figure*}

Comparing the TNBs studied here to previously characterized TNBs, we see that they are much more widely separated in terms of their Hill sphere occupation. Figure \ref{polar_contour} illustrates the mutual semi-major axis to Hill radius fraction for the ultra-wide TNBs characterized in this work and the systems listed in Grundy et al. (2011); the widest known TNB with a well-characterized orbit not in our sample is Teharonhiawako/Sawiskera at $a/R_H \sim 0.06$.

Grundy et al. (2011) showed that the previously known widely-separated, loosely-bound TNBs are only found on dynamically cold heliocentric orbits. Here we seek to confirm this relationship, and identify which dynamical population hosts the wide binaries; is it the cold Classical Kuiper Belt, or can a low-inclination extension of the hot Classical Kuiper Belt plausibly host the widely-separated binary systems? Figure \ref{helio_excite} illustrates the outer orbital excitation (given by $\sqrt{\sin(i_{\mbox{out}})^2 + e_{\mbox{out}}^2 }$) versus $a/R_H$. As in Grundy et al. (2011), we confirm that only dynamically cold populations host wide binaries. Further, systems with pericenters which suggest current or past encounters with Neptune have relatively low $a/R_H$, supporting the destructive nature of such encounters presented by Parker \& Kavelaars (2010). To quantify the difference between the binaries found in dynamically cold and hot populations, we compare the $a/R_H$ distributions of binaries falling into our CC classification with all other binaries. We find that these two samples are inconsistent with being drawn from the same $a/R_H$ distribution, with the KS test rejecting this hypothesis at greater than 99.9\% confidence. Therefore, the binaries which fall into our crude dynamical classification of the cold Classical Kuiper Belt have distinctly different characteristics than binaries hosted in other populations.

\begin{figure*}[t]
\begin{centering}
\includegraphics[width=\textwidth]{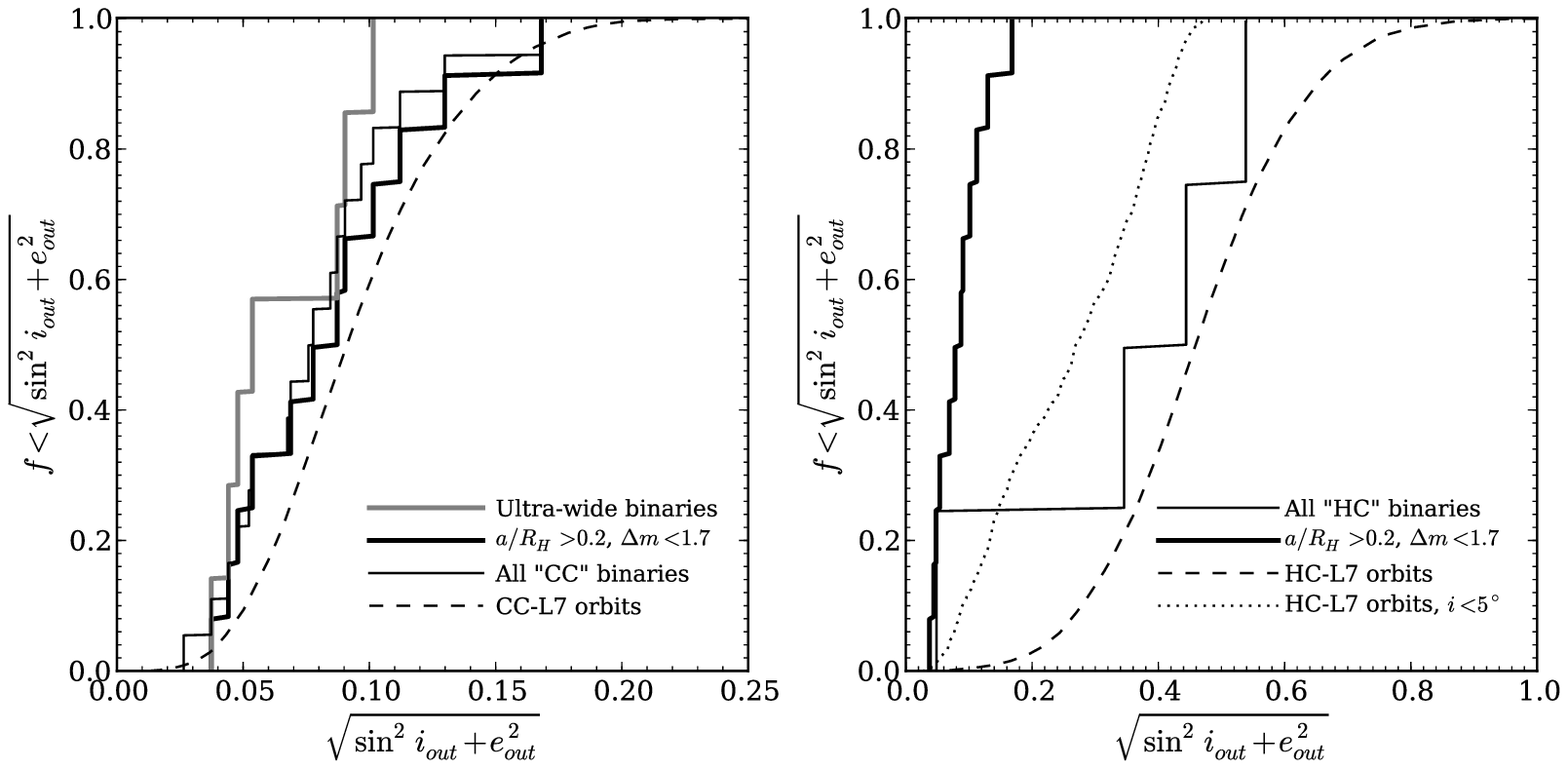} %%% fig 9
\caption{ Cumulative histograms of heliocentric orbital excitation for different binary sub-samples, compared to model distributions. Left panel: All CC binaries (heavy solid histogram), just the ultra-wide binaries (gray histogram), and all binaries with $a/R_H> 0.02$ and $\Delta m < 1.7$ (light black histogram), compared to the CC-L7 model distribution (dashed line). Right panel: All HC binaries (solid histogram) compared to the HC-L7 distribution (dashed histogram). We stress that this plot is comparing a heavily biased sample (binaries) with a de-biased model, and is only used for consistency checks.}
\label{helio_hist}
\end{centering}
\end{figure*}

To further clarify the host population of the wide binaries, we compare the outer orbital excitation distribution of the binaries with the distribution of the CC-L7 and HC-L7 orbital excitations. The distribution of orbital excitations for these binary samples and the L7 model populations are illustrated in Figure \ref{helio_hist}. Since most binaries considered here have been discovered in near-ecliptic surveys, there is significant bias against the detection of high inclination binaries. The bias is less significant when considering low inclination populations, so comparisons between the biased CC binary sample with the de-biased CC-L7 model population are reasonably fair. Comparing the CC binary sample to just the low inclination objects in the de-biased HC-L7 model population is also meaningful, as observational biases are not significant for the low-inclination extension of the HC-L7 population. We use the KS test to determine if we can reject the following two hypotheses: that the CC binaries are drawn from the CC-L7 orbital excitation distribution, or that the CC binaries are drawn from the excitation distribution of low-inclination ($i_{\mbox{out}} \leq 5^\circ$) HC-L7 orbits. Additionally, we determine whether just the seven ultra-wide binaries characterized in this work  (with no outer orbit constraints placed on them) could be drawn from the CC-L7 or  $i_{\mbox{out}} \leq 5^\circ$ HC-L7 excitation distributions, and whether \textit{all} binaries with $a/R_H > 0.02$ and $\Delta m < 1.7$ could be drawn from the same excitation distributions.

We rule out ($P < 0.001$) that the CC binaries are drawn from any subset of the HC-L7 excitation distribution, while it is plausible (KS test cannot reject at high confidence) that they are drawn from the CC-L7 excitation distribution. We also find that, without any prior cuts on heliocentric orbits, neither the seven ultra-wide binaries characterized in this work nor all binaries with $a/R_H > 0.02$ and $\Delta m < 1.7$ can be drawn from the low-inclination HC-L7 excitation distribution ($P < 0.001$ in both cases), while both can be plausibly drawn from the CC-L7 excitation distribution. This further confirms that wide binaries are intimately linked to the dynamically cold population of the Classical Kuiper Belt --- it cannot be that they are predominantly hosted by a low-inclination extension of the hot Classical Kuiper Belt.

\begin{table}
\centering
\begin{tabular}{ lll r}
\multicolumn{4}{c}{\bf Table 5}\\
\multicolumn{4}{c}{Kozai Oscillations} \\
\hline
Name & $e_{\mbox{max}}$ & $e_{\mbox{min}}$ & $q_{\mbox{min}}/R_p$$^*$\\
\hline
\hline
2001 QW$_{322}$ 	& $0.477^{+0.012}_{-0.006}$  	&  $0.342^{+0.017}_{-0.009}$  	&  $830^{+36}_{-29}$ \\
2003 UN$_{284}$  	& $0.45^{+0.04}_{-0.07}$    	&  $0.39^{+0.04}_{-0.08}$		&  $486^{+57}_{-42}$ \\
2006 BR$_{284}$ 			& $0.72^{+0.02}_{-0.02}$    	&  $0.263^{+0.009}_{-0.010}$  	& $155^{+15}_{-13}$ \\
2006 CH$_{69}$ 			& $0.94^{+0.01}_{-0.02}$  	&  $0.70^{+0.06}_{-0.08}$   	&  $31^{+9}_{-7}$  \\
\hline
2000 CF$_{105}$  	& $0.29^{+0.02}_{-0.02}$  	&  $0.28^{+0.02}_{-0.02}$ 	&  $739^{+29}_{-29}$ \\
2005 EO$_{304}$  	& $0.24^{+0.02}_{-0.03}$  	&  $0.22^{+0.02}_{-0.03}$  	&  $698^{+7}_{-6}$  \\
2006 JZ$_{81}$  			& $0.85^{+0.03}_{-0.02}$  	&   $0.84^{0.03}_{-0.02}$   	&  $82^{+10}_{-18}$ \\
\hline
\end{tabular}
{\footnotesize \\$^*$:  Primary radius $R_P$ assumes $\rho=1$ gram cm$^{-3}$.}
\end{table}

\subsubsection{Mutual inclination distribution}

The mutual inclination of a TNB system is one of the most challenging parameters to measure, as there is a mirror degeneracy in the pole solution which can only be broken after sufficient time has elapsed for the the observer's viewing geometry of the binary system has changed enough to discern the system's true orientation. However, the distribution of mutual inclinations and the ratio of prograde to retrograde orbits holds significant implications for formation scenarios (Schlichting \& Sari 2008b, Noll et al. 2008b) and for the ongoing evolution of the binary orbit (Fabrycky \& Tremaine 2007, Perets \& Naoz 2009). As illustrated in Figure \ref{polar_contour}, all seven systems characterized in this work now have non-degenerate pole solutions: four prograde and three retrograde. The prograde-to-retrograde ratio and its 95\% Poisson counting uncertainty for the ultra-wide TNBs is therefore $\sim1.33^{+4.55}_{-1.02}$. If we fold in the systems with non-degenerate pole solutions presented in this work and those in Grundy et al. (2011) which fall into our CC sub-sample and meet our near-equal mass criteria, we find that the ensemble prograde-to-retrograde ratio for dynamically cold, near-equal mass TNBs is $1.60^{+2.96}_{-0.99}$.

\begin{figure}
\begin{centering}
\includegraphics[width=0.5\textwidth]{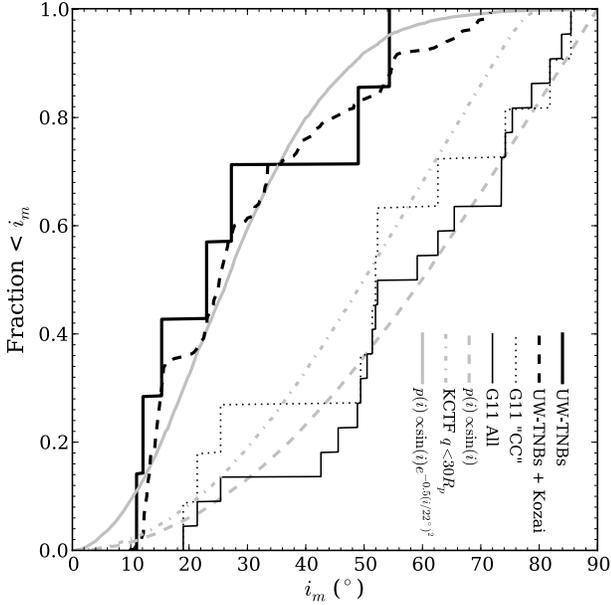} %%% fig 10
\caption{Cumulative mutual inclination distribution of ultra-wide binaries characterized in this work (heavy black histogram), tight binaries from literature (light black histogram), and tight CC binaries from literature (dotted black histogram). Retrograde inclinations are folded over onto the range of $0^\circ$---$90^\circ$. Dashed black line shows ultra-wide binaries' inclinations when smoothed over entire Kozai cycle of each object. Gray dashed line: uniform distribution --- $P(i)\propto sin(i)$. Grey dash-dotted line: uniform distribution modified by removing any orbits with Kozai cycles which drive $q_{\mbox{min}} \leq 30R_{p}$ as described in the text. Gray solid line: inclination distribution of form $P(i) \propto \sin(i) e^{-0.5(i/22^\circ)^2}$.}
\label{inc_dist}
\end{centering}
\end{figure}

The distribution of the mutual orbit inclinations of the ultra-wide binaries is illustrated in Figure \ref{inc_dist}, along with the inclination distribution of tightly-bound TNBs in literature, taken from the orbits compiled by Grundy et al. (2011). Due to the relative lack of high-inclination and excess of low-inclination wide binaries, we find it highly unlikely that the wide binaries' inclinations are drawn from the same inclination distribution as the tighter TNBs, with the Kolmogorov-Smirnov (KS) test rejecting this hypothesis at $>95$\% confidence. Additionally, we find that the tight binaries in literature are consistent with having their inclinations drawn from a uniform distribution ($P(i) \propto sin(i)$, with the KS test rejecting this hypothesis at only $\sim6$\% confidence), while the ultra-wide TNBs' inclinations are inconsistent with being drawn from the same uniform distribution (KS test rejecting this hypothesis at $\sim99$\% confidence). A simple way to understand the strength of this rejection is to note that in a uniform inclination distribution, 50\% of the inclinations are $>60^\circ$, while there are no systems in our sample with inclinations so high. Since the probability of randomly drawing an inclination $< 60^\circ$ is 0.5 each time, after sampling seven systems the probability of every system having inclination $< 60^\circ$ is the same as landing seven coin flips head-up in a row, $0.5^7 \simeq 0.008$.

When restricting the tightly-bound literature binaries to only those in the CC sample, the significance of the rejection of the wide binaries being drawn from the same distribution drops to only $>89$\% --- however, the dynamically cold binaries from literature remain consistent with a uniform distribution (KS test rejecting this hypothesis at only 40\% confidence). 

It should be noted that observational bias works against the detection of low inclination systems (with respect to the Ecliptic), as they  present edge-on geometry and their average projected separations are lower than high inclination systems. Thus, our relative lack of high inclination objects is not due to observational bias.

When considering all the inclinations the wide binaries reach during their Kozai cycle, we find that a $\sin(i)$ times a Gaussian distribution (frequently used to describe the inclination distribution of heliocentric Kuiper Belt orbits, eg., Brown 2001) centered at $i=0^\circ$ with width $\sigma\simeq22^\circ$ is compellingly similar to the Kozai-modified distribution. We find that widths between $10^\circ \leq \sigma \leq 50^\circ$ are consistent with the observed (non-Kozai) distribution within the 95\% confidence interval.

The difference in inclination distributions between both populations could either be due to cosmogonic variations inherent in different formation mechanisms, or due to evolutionary processes. Fabrycky \& Tremaine (2007) showed that tidal friction coupled with Kozai oscillations can cause wide, high-inclination systems to shrink and become circularized, creating a paucity of widely-separated, high-inclination binaries, and an excess of tight binaries near the critical inclinations for Kozai cycles ($40^\circ$ and $140^\circ$). This is referred to as the Kozai Cycles with Tidal Friction (KCTF) mechanism, and its plausibility as a significant evolutionary mechanism for binary minor planets was confirmed by Ragozzine (2009) and Perets \& Naoz (2009). While small numbers of objects and other complicating effects may prohibit the detection of an increase of tight binaries near the critical inclinations, the observed paucity of high-inclination, ultra-wide TNBs is suggestive of the KCTF effect in action.

We have performed a cursory test to compare the observed inclination distribution to the outcomes of KCTF. We determine the maximum eccentricity $e_{max}$ each binary system can reach for a grid of initial inclinations $i_0$ and arguments of pericenter $\omega_0$, assuming they begin with initial $e_m$ and $a_m$ equal to their present values. We then determine the fraction of phase space as a function of $i_0$ (assuming $i_0$ was initially uniformly distributed on a sphere) which do not lead to an $e_{max}$ which cause each system's pericenter to drop below a critical number of primary radii --- in other words, we require that $q_{min} \geq n \times R_{p}$. Ragozzine (2009) showed in full numerical simulations with reasonable assumptions that tidal dissipation became significant at $q \sim 20 R_{p}$ for the binary system Orcus/Vanth, which is more massive than the binaries studied here. However, the binary system 2006 CH$_{69}$ has Kozai cycles which take it to  $31^{+9}_{-7} R_{p}$, and presumably its existence suggests that such pericenter separations are stable for a significant fraction of the age of the solar system. As such, we chose the limit $q_{min} \geq 30 \times R_{p}$ for our cursory KCTF test,  and compare the resulting ``KCTF-modified'' inclination distributions to the observed ultra-wide TNB inclination distribution (illustrated in Figure \ref{inc_dist}). We find that the resulting distribution (and any with smaller $q_{min}$ cutoff) is ruled out at $97$\% confidence. 

We conclude that while the KCTF mechanism may have modified the orbits of some very high-inclination ultra-wide TNBs, it alone is not enough to explain the current mutual inclination distribution of these systems. To verify this, more comprehensive studies of the effects of KCTF on systems like those presented here are needed. If KCTF is not sufficient to explain the mutual inclination distribution of the ultra-wide TNBs, then we must consider cosmogonic effects. The primordial poles of the wide binaries may have preferred orientations orthogonal to the Ecliptic plane, suggesting formation in a very cold disk (Noll et al. 2008b). It should be noted that if the inclination distribution of the ultra-wide TNBs is non-uniform, and the binaries are subjected to small perturbations over their lifetimes (e.g., collisions), then the mutual inclination distribution will always tend to become \textit{more} random over time, approaching a uniform distribution. As such, the primordial inclination distribution would have to have favored low inclinations even more strongly than the current distribution does. We explore the effects of collisions on the mutual inclination distribution in an upcoming paper.

\section{ Albedos and Densities }

\begin{figure}[t]
\begin{centering}
\includegraphics[width=0.5\textwidth]{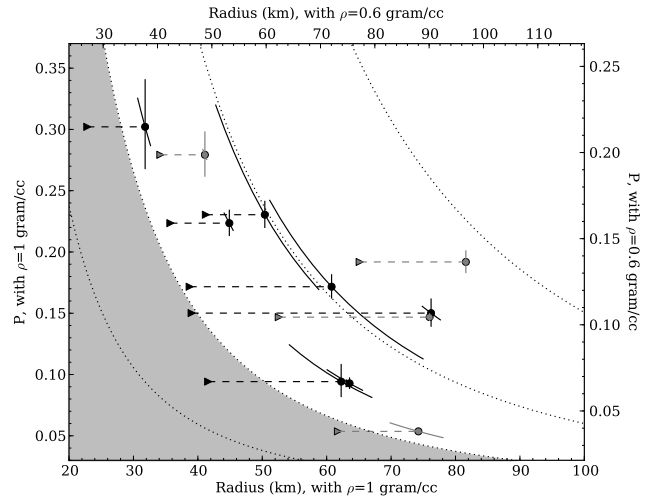} %%% fig 11
\caption{Albedos and radii for CC binary systems. Left/bottom axes show values assuming $\rho=1$ gram cm$^{-3}$, while top/right axes show the values assuming $\rho=0.6$ gram cm$^{-3}$. Circles mark primary radii, triangles connected with dashed line mark secondary radii. Black points represent binaries characterized in this work, gray points represent values derived from literature for the low-inclination classical TNBs 1998 WW$_{31}$, Borasisi/Pabu, Logos/Zoe and Teharonhiawako/Sawiskera. Dotted lines represent contours of constant \textit{H} magnitude. Gray region marks $H_r > 8$, likely unpopulated due to flux limits of current binary searches.}
\label{albedo}
\end{centering}
\end{figure}

Given the dynamically-derived masses and visible photometry for each system, we can explore the albedos and densities for the component bodies in these binaries. Without radiometric measurements to ascertain the albedo independently, the density and albedo remain degenerate. However, by assuming physically plausible values for the component densities, we can estimate the implied albedos for each object; the results of this exercise are illustrated in Figure \ref{albedo} and listed in Table 6. Generally, albedos for our sample of ultra-wide TNBs are found to be consistent with those measured radiometrically for larger solitary cold Classical Kuiper Belt objects (eg., Brucker et al. 2009), and range from 9\%---30\%, assuming $\rho = 1$ gram cm$^{-3}$ (6.4\%---21\% assuming $\rho = 0.6$ gram cm$^{-3}$). Figure \ref{albedo} also includes estimates of the albedos and radii of four other binaries from literature, and these systems were selected as members of the CC sample which had estimates of their $r$-band magnitude. The albedos of these four literature systems range from 5.4\%---28\% (with $\rho = 1$ gram cm$^{-3}$).

\begin{table}
\centering
\begin{tabular}{ lll r}
\multicolumn{4}{c}{\bf Table 6}\\
\multicolumn{4}{c}{Albedos and Primary Radii} \\
\multicolumn{4}{c}{(with $\rho=1$ gram cm$^{-3}$)}\\
\hline
Name & $P$ & $R_{p}$ & Note\\
& ($r'$) & (km) & \\
\hline
\hline
2000 CF$_{105}$ 	& 0.30$^{+0.04}_{-0.03}$ 		& 31.8$^{+0.6}_{-0.8}$ &$^1$\\
2001 QW$_{322}$ 	& 0.093$^{+0.010}_{-0.006}$ 	& 64$^{+1}_{-2}$&$^1$\\
2003 UN$_{284}$ 	& 0.09$^{+0.03}_{-0.01}$ 		& 62$^{+3}_{-5}$&$^1$\\
2005 EO$_{304}$ 	& 0.15$^{+0.01}_{-0.01}$	 	& 76.2$^{+1.0}_{-0.9}$&$^1$\\
2006 BR$_{284}$                     	& 0.22$^{+0.01}_{-0.01}$ 		& 44.9$^{+0.4}_{-0.5}$&$^1$\\
2006 JZ$_{81}$                       	& 0.17$^{+0.07}_{-0.06}$ 		& 61$^{+10}_{-6}$&$^1$\\
2006 CH$_{69}$                      	& 0.23$^{+0.09}_{-0.06}$ 		& 50$^{+6}_{-5}$&$^1$\\
\hline
1998 WW$_{31}$ 	& 0.054$^{+0.004}_{-0.004}$ 	& 74$^{+3}_{-3}$		&$^{1,3,7}$\\
Teharonhiawako 	& 0.147$^{+0.003}_{-0.003}$ 	& 76.0$^{+0.3}_{-0.3}$	&$^{1,4,7}$\\
Borasisi 			&0.192$^{+0.009}_{-0.009}$ 	& 81.6$^{+0.2}_{-0.2}$	&$^{2,5,7}$\\  %%% 1999 RZ$_{253}$
Logos 			& 0.28$^{+0.02}_{-0.02}$ 		& 41.1$^{+0.2}_{-0.2}$	&$^{2,6,7}$\\
\hline
\end{tabular}
\\{\footnotesize $^1$: Adopting $m_\odot = -26.93$. $^2$: Adopting $m_\odot = -27.12$. $^3$: Photometry from Veillet et al. (2002). $^4$: Photometry from Benecchi et al. (2009). $^5$: Photometry from Delsanti et al. (2001). $^6$: Photometry from Jewitt \& Luu (2001). $^{7}$: Mass from Grundy et al. (2011).}
\end{table}

The apparent trend of increasing albedo with decreasing radius visible in Figure \ref{albedo} is due to a selection effect: in a flux-limited survey, any physically small object detected must have a high albedo. The apparent trend here is consistent with a flux limit somewhat less than $H_r \simeq 8$. We note that, in general, the observations which discovered the primary of a given binary system were not the observations which discovered the binary nature of the system. Thus, this flux limit seems to apply to the primary absolute magnitude, and the fact that secondary absolute magnitudes scatter across the $H_r = 8$ line reflects deeper follow-up observations identifying the secondaries.

\begin{figure*}
\begin{centering}
\includegraphics[width=\textwidth]{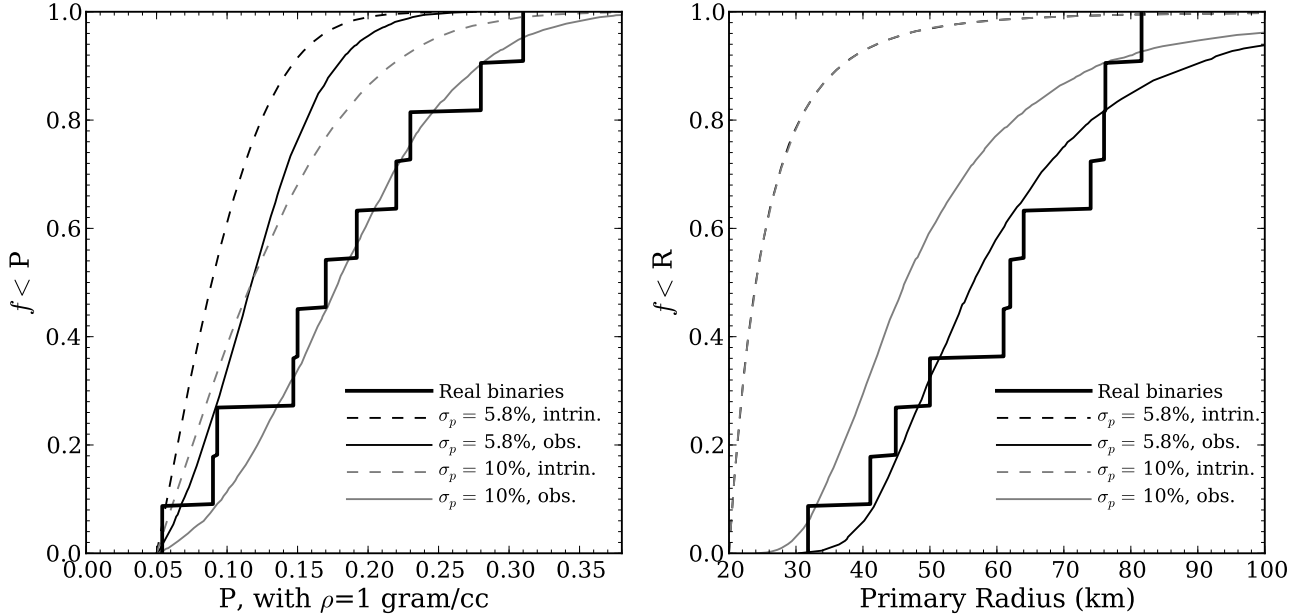} %%% fig 12
\caption{Left panel: comparison of binary albedo distribution (heavy histogram) with model distributions, assuming an albedo distribution with form of Eqn. \ref{albedo_eqn}, a power-law size distribution with slope $q=4.8$, and binary discovery survey flux limit of $H_r = 8$. Intrinsic albedo distributions shown by dashed lines, while synthetic ``observed'' distributions shown by solid lines. Right panel: Same binaries and albedo distributions, but showing resulting radius distributions.}
\label{albedo_dist}
\end{centering}
\end{figure*}

The lack of a strong group of small, high-albedo binaries along with the lack of large, high-albedo binaries suggest that relatively low albedos may be more common than high albedos in the cold Classical Kuiper Belt. All seven binaries in Figure \ref{albedo} with primary radius greater than 55 km have nominal albedos less than 20\%, while all four binaries with primary radius smaller than 55 km have nominal albedos greater than 20\%. Given the steepness of the size distribution of low-inclination TNOs in this size range ($q\sim4.8$, Fraser \& Kavelaars 2009), we would expect there to be roughly nine times as many objects in the radius radius range from 30---55 km than at all radii larger than 55 km. Since the sample contains only roughly 0.57 times as many binaries in the smaller size range, we posit that the rest of the expected small binaries are missing due to having low albedos, making them invisible to the flux limits of the surveys which discovered the binary systems in our sample.

We adopt an ansatz albedo distribution of a Gaussian centered at $p=0.05$, and clipped such that $p > 0.05$ (comparable to the lowest measured albedo in our sample):

\begin{equation}\label{albedo_eqn}
P(p) \propto \left\{ \begin{array}{ll} 
e^{-0.5 ( \frac{ p - 0.05 } { \sigma_p } )^2 } 	&: p > 0.05 \\
0										&: p \leq 0.05 \end{array} \right.
\end{equation}

Additionally, we adopt a size distribution with slope $q=4.8$, and estimate the flux limit at discovery to be $H_r = 8$. When we draw a radius from this size distribution, we assign it an albedo from our Gaussian albedo distribution and determine if it is bright enough to have been observed by our synthetic survey (brighter than $H_r = 8$). We then compare the properties of these synthetic ``observed'' systems with the systems which were actually observed; we vary the width $\sigma_p$ of the albedo distribution until the KS test can rule out that either the distribution of real radii or the distribution of real albedos are drawn from the synthetic ``observed'' distributions. We find that at 95\% confidence, the observed range of real albedos implies that the albedo distribution must have a width $\sigma_p \gtrsim 0.058$ or there would be too few high albedo detections. Additionally, given the observed real primary radius distribution, the albedo distribution must have a width $\sigma_p \lesssim 0.1$ or there would be too many small objects discovered. Figure \ref{albedo_dist} illustrates the distribution of albedos and primary radii of the real binaries in comparison to the synthetic ``observed'' distributions of these parameters.

The lack of a strong group of small, high-albedo binaries may also be explained in the case of a more uniform albedo distribution by positing that the binary fraction decreases drastically for decreasing radii, similar to the prediction of Nesvorn\'{y} et al. (2011). However, with the addition of a varying binary fraction with size as a new degree of freedom, the current sample size is not sufficient to quantitatively constrain such behavior at this time. We note that given the large range of albedos observed in this population, any sharp features in the trend of binary fraction with radius will be blurred if considering only the absolute magnitude of systems (as done in Nesvorn\'{y} et al. 2011), and any such features will be much more evident when radii derived from mass measurements or radiometric measurements are used in place of absolute magnitudes.

\section{Discussion}

\subsection{Formation Mechanisms and Implications}

Since the discovery of the first TNBs, a number of possible formation pathways have been posited. In the following discussion, we consider those mechanisms most likely to form widely-separated, near-equal mass systems, and compare the predicted outcomes of each of these pathways to our observed sample.

\subsubsection{$L_2 s$ and $L_3$ mechanisms}

Originally described in Goldreich et al. (2002), these mechanisms were further investigated by Noll et al. (2008b) and Schlichting \& Sari (2008 a \& b). The $L_2 s$ mechanism posits that binaries are captured when two passing solitary objects can disperse some excess kinetic energy into a sea of smaller bodies and become bound, while the $L_3$ mechanism instead sends the excess kinetic energy away through scattering a third large body. 

Schlichting \& Sari (2008b) show that models like $L_2 s$, which rely on a smooth dissipation process to capture binaries, will dominate the binary formation rate only in conditions where the relative velocity between planetesimals $v$ is much less than the Hill velocity, $v_H = 2\pi R_H / T_{\mbox{out}}$, where $T_{\mbox{out}}$ is the heliocentric orbital period. They also show that under these conditions, the binary mutual inclinations will be dominantly retrograde, predicting a prograde-to-retrograde ratio $\lesssim 0.03$. The measured wide binary inclinations exclude such an extreme ratio of prograde to retrograde systems. Therefore, we can rule out this mechanism for forming wide binaries, unless an intervening dynamical process can be invoked to re-orient a large number of binary systems or preferentially destroy retrograde binary systems. We estimate that starting from a primordial prograde-to-retrograde ratio of 0.03, at least 22\% of wide binary systems would have to be re-oriented in order to not be ruled out at greater than 95\% confidence by the current observed prograde-to-retrograde ratio.

Under more energetic conditions, where the relative velocity between planetesimals exceeds $v_H$, Schlichting \& Sari (2008b) show that three-body interactions ($L_3$ models) will dominate the binary formation rate. In this regime, they find that roughly equal numbers of prograde and retrograde systems are formed, consistent with the observed distribution of inclinations. However, they also show that only systems with separations of order $s \lesssim R_H (v_H/v)^2$ tend to survive the formation phase, and binary formation rates drop dramatically as $v$ increases. As our observed binary systems have separations of order $0.08-0.23R_H$, the velocity dispersion in the primordial disk could not have exceeded 2---4 times $v_H$ if this formation mechanism applied. Additionally, Noll et al. (2008b) suggest that formation in a dynamically cold disk ($v < v_H$) should produce aligned orbit poles. Since the wide binaries seem to prefer low mutual inclinations, this argues for formation in a dynamically cold disk, but not so cold as to allow $L_2 s$ to dominate and produce a large fraction of retrograde systems.

Together, the widely-separated components, lack of clear preference for retrograde orbits, but apparent preference for low mutual inclinations all point toward the velocity dispersion being approximately equal to the Hill velocity. This represents a fine-tuning problem (Noll et al. 2008b), for there is no clear \textit{a priori} reason to expect that $v \sim v_H$. Additionally, it is not clear whether the balance between the $L_3$ and $L_2 s$ mechanisms at $v \sim v_H$ would simultaneously produce widely separated binaries, aligned poles, \textit{and} roughly equal numbers of prograde and retrograde orbits.

\subsubsection{Exchange Reactions and Chaos-Assisted Capture}

Funato et al. (2004) suggest that multiple exchange reactions (where one object in a binary system is swapped for a passerby) can produce very widely separated binary systems. However, systems as widely separated as those in our sample formed through exchange reactions all have very high eccentricities ($e_m \gtrsim 0.9$, see Figure \ref{polar_contour}). The systems 2006 CH$_{69}$ and 2006 JZ$_{81}$ have present eccentricity values consistent with these predictions, but all other systems are presently inconsistent with such high eccentricities. Several systems are subject to large oscillations of their inclination and eccentricity due to Kozai cycles, but these oscillations do not carry them to eccentricities as high as predicted by exchange reactions. Thus, it seems unlikely that this mechanism dominated binary formation.

Astakhov et al. (2005) simulated the effect of chaotic transient binaries on stable binary formation. They found that two objects temporarily caught in their mutual chaotic layer could become stabilized by dynamical friction due to a sea of small objects --- effectively adding an enhancement to the $L_2 s$ mechanism due to transient, chaotic orbits. This mechanism is referred to as Chaos-Assisted Capture. They find mutual eccentricities spanning the range of those observed in our sample, but the separations they find for binaries formed by Chaos-Assisted Capture do not extend to as high as those found for the ultra-wide TNBs (see Figure \ref{polar_contour}). Additionally, Schlichting \& Sari (2008a) argue that the enhancement due to these transient captures is not significant, and that formation should proceed as they found for the $L_2 s$ and $L_3$ mechanisms. We conclude that it is unlikely that this mechanism dominated the ultra-wide TNB formation rate.

\subsubsection{Gravitational collapse}

Recently, another mechanism has been proposed to form Kuiper Belt binaries. Operating with the framework of planetesimal formation through rapid gravitational collapse in a turbulent disk, Nesvorn\'y et al. (2010) suggest that binaries may form as a cloud of cm-scale particles collapses and fragments. This model produces binaries very efficiently, and their properties can vary widely. Mass ratios of order unity are produced, and semi-major axes from $10^3 - 10^5$ km are produced for systems with primary radii ranging from tens to hundreds of kilometers. Broad ranges of inclination and eccentricity can be produced for all semi-major axes. Additionally, this mechanism has the attractive feature of producing a natural explanation for the correlated colors of binary components (Benecchi et al. 2009), in contrast to the broad range of colors exhibited between different binary systems.

\begin{figure}
\begin{centering}
\includegraphics[width=0.5\textwidth]{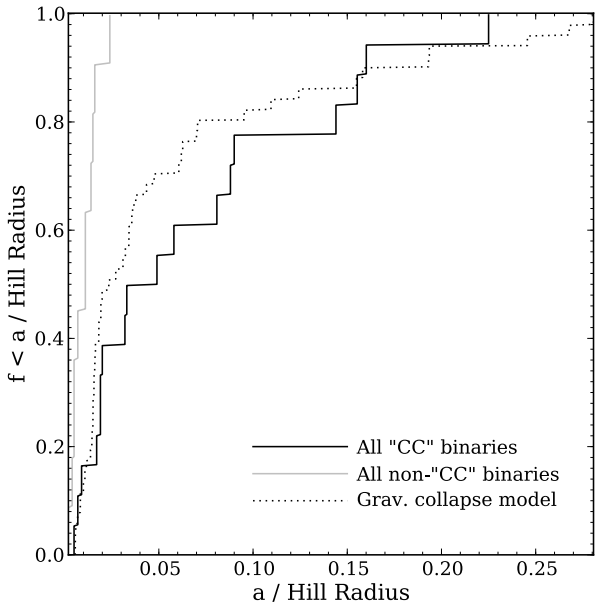} %%% fig 13
\caption{Cumulative histogram of $a/R_H$ for CC TNBs (solid histogram) and all other TNBs with known orbits (dashed histogram). Also shown is results of gravitational collapse binary formation simulations by Nesvorn\'{y} et al. (2010), selecting simulations with $\Omega=0.5-1.0\Omega_{circ}$ (initial clump rotation), $f^*=10-30$ (cross-section modifier), and final mass ratio $<$ 10.}
\label{aRH_hist}
\end{centering}
\end{figure}

We compare the results of a subset of these simulations to our observed mutual orbits. We select only those simulations which produce binaries with final mass ratio $<$ 10, and with initial particle-swarm rotation $\Omega = 0.5-1.0 \Omega_{circ}$ (where $\Omega_{circ} =  R^{-\frac{3}{2}} \sqrt{G M}$, and $M$ and $R$ are the total initial mass and radius of the swarm) and collisional cross-section enhancements (to account for the lower resolution of the simulation compared to reality) $f^* = 10-30$. See Nesvorn\'{y} et al. (2010) for a more thorough description of these parameters and their importance. Figure \ref{polar_contour} illustrates the mutual eccentricity and $a/R_H$ for all the synthetic orbits formed by this mechanism which meets these criteria, and in general we find that they appear to mimic the distribution of real orbits surprisingly well. Figure \ref{aRH_hist} illustrates the distribution of $a/R_H$ for the synthetic orbits compared to the CC binaries and all the other binaries. We find that the $a/R_H$ distribution of the CC binaries is statistically indistinguishable from the synthetic orbit distribution generated by gravitational collapse, while we rule out that the other binary populations were drawn from the same distribution as the synthetic orbits at a high level of confidence.

This agreement is encouraging, but it remains preliminary. The initial conditions for binary formation by gravitational collapse is still highly uncertain, and it remains unclear if this mechanism can produce a large number of retrograde binaries or if the binaries it produces will exhibit a preference for low mutual inclinations. Additionally, while the $a/R_H$ distributions appear to agree, there are likely a number of biases in the current observed distribution and any conclusions drawn from it must be taken with caution. The binary fraction is measured to be $\sim30$\% in the cold Classical Kuiper Belt (Noll et al. 2008a) at the limit of HST resolution, while the fraction of ultra-wide systems is approximately 1.5\% (Lin et al. 2010); therefore, the ultra-wide binaries make up approximately 5\% of the current CC binary population. However, comparing the systems with measured orbits, we see that the ultra-wide systems are over-represented, with our sample alone making up $\sim39$\% of all the CC binaries with measured mutual orbits. Taking this into account, it appears that the Nesvorn\'{y} et al. (2010) model actually over-produces wide binaries compared to the present population.

There is additional concern since the current $a/R_H$ distribution may not represent the primordial distribution; Petit \& Mousis (2004) showed that these wide binary systems were likely much more common in the past, and may have been reduced in number due to collisional disruption. We discuss this problem in more detail in the following section.

\subsection{Susceptibility to Disruption}

Because of their low binding energy, ultra-wide TNBs are very susceptible to disruption by a variety of perturbations. Parker \& Kavelaars (2010) simulated the transplantation of the Cold Classical Kuiper Belt from closer to the Sun via Neptune scattering as suggested by Levison et al. (2008), and found that wide binaries are very efficiently destroyed by close encounters with Neptune before they can be implanted in the current Kuiper Belt. The binaries we characterize in this work are all wide enough to be easily stripped by this process, and comparing the $a/R_H$ and eccentricity of our systems to the results of Parker \& Kavelaars (2010), we find destruction probabilities ranging from at least 75\% for 2006 BR$_{284}$ to over 98\% for 2001 QW$_{322}$. Furthermore, such interactions would likely have randomized the orbit poles of the surviving wide binaries, not leaving behind the aligned poles we see today.

Taken together, we estimate that if our sample represents a population which has been subjected to disruption by Neptune scattering, the initial population of ultra-wide TNBs would have to have been roughly 13 times larger than the current population. If we take the lower limit of the current wide binary fraction of the Cold Classical Kuiper Belt estimated by Lin et al. (2010) at 1.5\%, this indicates that the primordial wide binary fraction have to exceed 20\% to leave enough wide binaries surviving post-Neptune scattering. Correcting the estimate of Kern \& Elliot (2006) to apply to only the Cold Classical Kuiper Belt (using the same approach as Lin et al. 2010) results in an estimate closer to 5\% for the current wide binary fraction, which would imply a primordial fraction of wide binaries in excess of 65\%. These estimates assume no other processes disrupted binaries in the intervening time between implantation in the Kuiper Belt and the present day. A primordial wide binary fraction of 20\% is comparable to the entire current binary fraction in the Cold Classical Kuiper Belt.

As mentioned previously, the Nesvorn\'{y} et al. (2010) simulations appear to over-produce wide binaries with respect to today's population; synthetic binaries as wide or wider than the ultra-wide TNBs characterized in this work created by the Nesvorn\'{y} et al. (2010) model represent roughly 20\% of the systems produced by their simulations. Therefore,  if the primordial binary fraction was $\sim100$\%, the ultra-wide binary population produced in these simulations would be roughly sufficient to leave a remnant population similar to the observed population after implantation in the cold Classical Kuiper Belt through Neptune scattering if no additional processes disrupted these systems afterward. However, since a much larger fraction of the tighter binaries would survive the implantation process, we would expect a significantly larger remnant population of tight binaries if the primordial binary fraction was so high. Later processes (such as collisions) could modify the binary fractions over the age of the solar system, but generally wide binaries are more susceptible to stripping processes and the relative fraction of wide to tight binaries will tend to decrease over time. Therefore, barring a formation mechanism which much more strongly favors wide separations, we find that the ultra-wide binary orbits confirm the findings of Parker \& Kavelaars (2010) that it is unlikely for the cold Classical Kuiper Belt to have experienced a period of scattering encounters with Neptune and likely formed in situ. 

%and at present we are aware of no binary formation scenario which produces such large numbers of wide binaries.

In addition to being susceptible to disruption by tidal effects, collisions are effective at unbinding these systems. Petit \& Mousis (2004) estimated the mean lifetimes for wide binaries given an estimate of the number of impactors in the current Kuiper Belt capable of unbinding these systems in an impact. They found that binaries like 2001 QW$_{322}$ had mean lifetimes much less than the age of the solar system, and required a primordial population of wide binaries at least seven times larger than the extant population to explain the continued existence of several such systems.

Following Petit \& Mousis (2004) and A.H. Parker \& Kavelaars (2011, in submitted), we estimate the mean lifetime of a binary on a circular mutual orbit due to collisional unbinding by an impact on the secondary to be:

\begin{equation}
\begin{split}
\tau_s^{-1} = N_0 R_0^{q-1} P_i R_s ^ { (3-q) } \mbox{\hspace{2 cm}} \\ \times \left( \frac{0.62}{V_i} \right)^{(1-q)/3}  \left( \frac{ G M_{sys} } { a_m } \right) ^ { (1-q)/6 },
\end{split}
\end{equation}

\noindent where $N_0$ is the number of impactors larger than $R_0$, drawn from a power-law size distribution with slope $q$, $V_i$ is the mean impact velocity, and $P_i$ is the intrinsic collisional probability. For the following discussion, we adopt the same values as Petit \& Mousis (2004) for $P_i = 1.3 \times 10^{-21}$ km$^{-2}$ yr$^{-1}$, and $V_i = 1$ km s$^{-1}$, which were based upon Farinella et al. (2000).

This calculation can be somewhat refined by noting that collisions onto the primary can also contribute to unbinding the system, and when folding in this second decay channel the system's mean lifetime becomes half the harmonic mean of the individual collisional lifetimes of the primary and the secondary,

\begin{equation} \label{tau}
\begin{split}
\tau^{-1} = N_0 R_0^{q-1} P_i \left( \frac{0.62}{V_i} \right)^{(1-q)/3}  \left( \frac{ G M_{sys} } { a_m } \right) ^ { (1-q)/6 } \\ \times \left(  R_s^ { (3-q) } +  R_p^ { (3-q)} \right).\mbox{\hspace{3 cm}}
\end{split}
\end{equation}

For equal-mass systems like 2001 QW$_{322}$, this effectively halves the estimated mean lifetime.

For an impactor population, we take the estimated total population of TNOs in the CFEPS L7 model which cross through the cold Classical Kuiper Belt region and extrapolate to small sizes. To do this, we first estimate the number of objects at $H_g = 10$ (roughly the magnitude of the putative break in the luminosity function of the Kuiper Belt) by extrapolating from the limits of the CFEPS survey ($H_g \sim 8.5$) with a luminosity function power-law slope of 0.76 (e.g., Fraser \& Kavelaars 2009). We estimate that there are of order $N_0 \sim 800,000$ objects brighter than $H_g = 10$ in which pass through the cold Classical Kuiper Belt. Translating this to a radius with a geometric albedo of 0.1, we adopt $R_0 \simeq 26$ km as our break and normalizing radius for the small-object size distribution. 

A number of groups (Bernstein et al. (2004), Fraser \& Kavelaars (2009), Fuentes et al. (2009)) have found that the luminosity function of the Kuiper Belt breaks to a shallower slope of approximately $\alpha \simeq 0.2$, translating to a power-law slope of $q\simeq2$. However, Fraser (2009) showed that the size distribution likely steepens again after a ``divot,'' and the true small object size distribution slope may be somewhat steeper than $q=2$. To capture this uncertainty in the behavior of the size distribution below the break, we estimate the collisional lifetimes of the ultra-wide TNBs using both a shallow slope of $q=2$ and a collisional-equilibrium slope of $q=3.5$. This steeper slope is comparable to the estimate of the size distribution slope at small sizes set by the putative detection of a single sub-km TNO by stellar occultation presented in Schlichting et al. (2009).

\begin{table}
\centering
\begin{tabular}{ lcccc }
\multicolumn{5}{c}{\bf Table 7}\\
\multicolumn{5}{c}{Collisional Lifetimes} \\
\hline
Name & \multicolumn{2}{c}{$q=2$} & \multicolumn{2}{c}{$q=3.5$}\\
& $\tau$ (yr) & $n^{*}$  &  $\tau$ (yr) & $n^{*}$ \\
\hline
\hline
2000 CF$_{105}$ & 5.0$\times10^{10}$ & 1 & 1.0$\times10^{9}$ & 52 \\
2001 QW$_{322}$ & 2.6$\times10^{10}$ & 1 & 2.7$\times10^{9}$ & 4 \\
2003 UN$_{284}$ & 3.3$\times10^{10}$ & 1 & 2.5$\times10^{9}$ & 5 \\
2005 EO$_{304}$ & 3.1$\times10^{10}$ & 1 & 2.9$\times10^{9}$ & 4 \\
2006 BR$_{284}$ & 4.2$\times10^{10}$ & 1 & 2.2$\times10^{9}$ & 6 \\
2006 JZ$_{81}$ & 3.7$\times10^{10}$ & 1 & 3.0$\times10^{9}$ & 4 \\
2006 CH$_{69}$ & 3.9$\times10^{10}$ & 1 & 2.7$\times10^{9}$ & 5 \\
\hline
\end{tabular}

$^{*}$: Estimate of primordial number of wide binaries with same system properties in order to leave one system surviving today, from $n = e^{( t / \tau )}$ with $t=4\times10^9$ years.
\end{table}

Collisional lifetimes from Eqn. \ref{tau} for each binary system are presented in Table 7. This table also contains an estimate of the required initial number of binaries $n$ to leave behind one system with a given set of properties after $t = 4\times10^9$ years of collisional bombardment, given $n = e^{( t / \tau )}$. For the shallow small-object size distribution slope $q=2$, all the binary systems characterized in this work have collisional lifetimes much in excess of the age of the solar system, and if this slope represents the size distribution then the binary population we see today can be taken as representative of the primordial population. However, in the case of the steeper slope $q=3.5$, we see that all the systems have lifetimes less than the age of the solar system, and each binary system represents a member of a decayed initial population ranging from $4-52$ times larger than the current population (the average required primordial wide binary population for this size distribution is $\sim11$ times larger than the extant population). 

We conclude that if the power-law size distribution slope at small sizes remains shallow, then the extant wide binary population can be taken as representative of the primordial wide binary population. If this is the case, then the preliminary agreement between the predictions of binary formation by gravitational collapse and the observed orbital distribution of wide TNBs remains valid when considering the primordial wide binary orbital distribution.

The estimates of collisional lifetimes presented here are very simple analytical estimates, and ignore some important effects such as the eccentricity of the binary system, mass loss during impacts, and solar tides. We have refined these calculations and performed extensive numerical simulations of binary disruption by collisions, and will present the results of this work in an upcoming publication.

The system 2000 CF$_{105}$ has by far the shortest collisional lifetime, due to its wide separation and very small component sizes; in fact, it is the second-widest known TNB (with respect to $a/R_H$) and its components have the lowest mass of any TNO currently measured. It can be unbound by the smallest impactors of all the binary systems considered here, and is therefore most sensitive to the impactor population at small sizes. The frequency of systems like 2000 CF$_{105}$ in the current Kuiper Belt is therefore of considerable interest, as they act as powerful probes of the collisional environment. Future surveys will likely identify many such systems, and we discuss the prospects for these surveys in the next section.

\subsection{Processes in the primordial disk}

In the previous sections, we have found that the inclination distribution of the wide binaries indicates a very cold dynamical environment at the time of formation, but formation of the binaries post-accretion through the $L_2 s$ pathway --- which should dominate in such conditions --- can be ruled out due to a lack of preference for retrograde orientations. We also found that the cold Classical Kuiper Belt was never subjected to a period of close encounters with Neptune, and therefore likely formed in situ. Finally, the properties of mutual orbits produced by current simulations of gravitational collapse are consistent with the observed properties of binaries in the cold Classical Kuiper Belt. 

Together, all these results suggest that formation of the cold Classical Kuiper Belt occurred in situ, with objects forming via rapid collapse of small particles directly into large ($10-100$ km) planetesimals through primary accretion driven by turbulent concentration of solids in the protoplanetary disk (eg., Johansen et al. 2007, Cuzzi et al. 2010).  This process offers a mechanism for forming binaries such as are observed in this population, and can feasibly form the cold Classical Kuiper Belt in situ (Cuzzi et al. 2010) without requiring that a large fraction of the primordial mass in planetesimals be subsequently removed through either significant collisional grinding or dynamical processes (both of which can be destructive to the binary population), even when the density of the primordial disk is required to drop sharply beyond 30 AU to halt the early migration of Neptune at its present distance. Additionally, since this process produces no massive sea of small planetesimals (most planetesimals being ``born big''), the $L_2 s$ mechanism will naturally be ineffective, thus preventing it from forming a dominantly retrograde binary population in the dynamically cold environment indicated by the binaries' low mutual inclinations. 

The mechanism of binary formation by gravitational collapse should be extensively explored. If it can be shown to produce roughly equal numbers of prograde and retrograde binaries while maintaining a preference for low mutual inclinations, then this should be taken as strong support for the formation of the cold Classical Kuiper Belt in situ by rapid primary accretion driven by turbulent concentration and gravitational collapse.

\subsection{Characterizing wide binaries with next-generation surveys}

%%% mean: 0.67"
%%% 25th percentile: 0.44"
%%% 75th percentile: 0.81"
%%% Population of Cold Classicals from CFEPS that are bright enough to be detected by LSST: limit ~8.5, roughly 17,000 objects.
%%% Assume same Boltzmann inclination distribution as found earlier (mu=0, sigma=23)
%%% Expect to visit each field of order 100-200 times (g and r).

While the current sample of known ultra-wide TNBs remains fairly small, future wide-area surveys like the Large Synoptic Survey Telescope (LSST) and Pan-STARRS will be quite capable of identifying such systems. The projected LSST $g$-band single-visit depth is $\sim25$, which translates to $H_{g} \simeq 8.5$ for a geocentric distance of 45 AU. At this depth, there are roughly 17,000 objects in the CC-L7 component of the CFEPS synthetic model of the Kuiper Belt. At a minimum, the results of Lin et al. (2010) suggest that at least 1.5\% of these objects are ultra-wide TNBs with characteristics like those in our sample, and we therefore expect over 250 such systems will be detectable by the LSST pipeline. 

\begin{figure}
\begin{centering}
\includegraphics[width=0.5\textwidth]{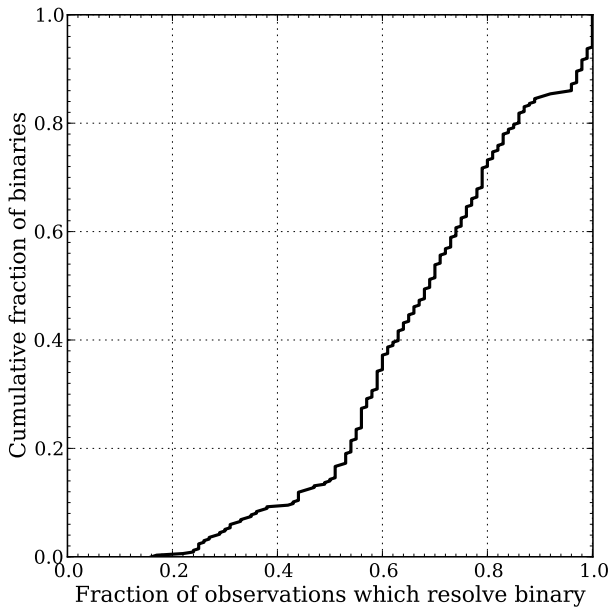}
\caption{Results of observing clones of our sample of ultra-wide TNBs through simulated LSST seeing, assuming observations in the $r$-band taken over the course of the 10-year LSST survey lifetime. An observation is said to be ``resolved'' when the full-width at half-max of the seeing during that observations is smaller than the projected binary separation.}
\label{LSST}
\end{centering}
\end{figure}

Some fraction of the observations of each of these TNBs will likely be resolved, depending on the binary separation and the seeing at the time of each observation. In order to estimate this fraction of resolved observations, we have used clones of each binary in our sample to determine the distribution of on-sky separations that can be expected over the ten-year baseline of the LSST survey. We move each clone to a  random starting point on both its heliocentric orbit and mutual orbits, then propagated its motion forward for ten years while sampling its on-sky separation at 100 ``observation epochs'' over than ten-year period. At every ``observation epoch'' we determined if the system was resolved or not based on its on-sky separation and an image quality drawn from the measured $r$-band distribution of seeing at the LSST site (LSST Science Book Version 2.0, Section 2.2), with median seeing of $\sim0\arcsec.6$. Figure \ref{LSST} illustrates the cumulative fraction $f_B$ of binary systems versus the minimum fraction of observations which resolve those systems.

The minimum fraction of resolved observations for any system we simulated was $\sim$18\%, and the maximum was 100\%. If we take the planned baseline number of 230 visits per field in the $r-$band to represent the number of observations likely to be made of each binary, then we would expect $\sim$40---230 resolved observations of each binary system. This indicates that virtually \textit{all} 250 TNB systems with properties similar to those in our sample likely to be observed by the LSST will have more resolved observation epochs than the most-observed binary in our current data set (2001 QW$_{322}$, with 35 observed epochs). As such, not only will LSST \textit{detect} most of these binary systems, but it will measure their mutual orbit properties at least as well as we have managed to do with the sample presented in this paper.

With such a large catalog of well-characterized binary orbits, all drawn from the same well-calibrated survey, unprecedented avenues of investigation will be opened. Identifying trends of binary properties with heliocentric orbit may provide further evidence for different origins and histories of the sub-components of the Kuiper Belt, and will provide new constraints on the dynamical history of the outer Solar System (Parker \& Kavelaars 2010, Murray-Clay \& Schlichting 2011), trends of binary fraction with component sizes will constrain the extent of collisional grinding in the early Kuiper Belt (Nesvorn\'{y} et al. 2011), and color---albedo trends may be identified and used to constrain the surface composition of these objects.

\section{Summary}

We have presented the first-ever well-characterized mutual orbits for a sample of seven ultra-wide Trans-Neptunian Binaries. These orbits range over eccentricities from $0.2-0.9$, and have semi-major axis to Hill Radius ($a/R_H$) fractions ranging from $0.08-0.22$. We find that their properties are distinct from other binary populations, with the following highlights:

\begin{enumerate}

\item The outer orbits of all widely-separated binaries ($a/R_H > 0.02$) with near-equal mass components ($\Delta m < 1.7$, or $M_p / M_s < 10$) are distributed consistently with being drawn from the ``stirred'' and ``kernel'' components of the CFEPS L7 synthetic model of the Kuiper Belt, and inconsistent with being drawn from the low-inclination subset of the ``hot'' component. This confirms that only the dynamically cold components of the Classical Kuiper Belt are host to wide binaries.

\item Two ultra-wide TNBs have very high mutual eccentricities ($e_m \simeq$ 0.84 and 0.9) consistent with the predictions of binary formation by exchange reactions (Funato et al. 2004), but the rest are inconsistent with the extreme eccentricities predicted by this mechanism.

\item Ultra-wide TNBs have roughly equal numbers of prograde and retrograde orbits, and the observed ratio is inconsistent with the predicted preference for retrograde orbits if the $L_2 s$ binary formation mechanism dominated unless over $\sim22$\% of wide binary systems have switched orientations over their lifetimes.

\item Ultra-wide TNBs have a statistically significantly different mutual inclination distribution compared to tighter binaries in literature, and their inclinations also cannot have been drawn from a uniform sphere ( $P(i) \propto sin(i)$ ) due to a lack of detections of high-mutual inclination wide binaries. This paucity of wide binaries at high inclinations and preference for low inclinations cannot be explained by the KCTF mechanism alone, and suggests a primordial preference for low-inclination mutual orbits. This suggests formation in a very dynamically cold disk. %, but more comprehensive analysis is needed.

\item The wide separations of these systems indicates that if binary formation was dominated by the $L_3$ mechanism, the velocity dispersion in the primordial disk must have been less than a few times the Hill velocity. A low velocity dispersion supports the observed preference for low mutual inclinations. However, the velocity dispersion could not have been much below the Hill velocity or the $L_2 s$ mechanism would have dominated and the ultra-wide binaries should exhibit a strong preference for retrograde orbits. Thus, under the assumption that these are the only two efficient formation pathways, it appears that $v \sim v_H$. Further modeling of these mechanisms is required to determine whether a balance can be struck between them where widely-separated binaries can be formed with aligned poles and roughly equal numbers of prograde and retrograde orientations.

\item Current simulations of the alternative formation mechanism of gravitational collapse and fragmentation (Nesvorn\'{y} et al. 2010) create orbital distributions similar to the observed present-day distributions. These simulations tend to over-produce wide systems, which leaves some room for post-formation disruption of ultra-wide binaries. However, it remains unclear if this mechanism can produce a large number of retrograde systems.

\item Assuming realistic densities, the implied albedos for the ultra-wide TNBs range over $\sim 0.09 - 0.30$, consistent with estimates of the albedos of solitary Cold Classical objects (eg., Brucker et al. 2009). The distribution of observed albedos with primary radius for all dynamically-cold TNBs (seven from this work and four others from literature) suggests that albedos at the lower end of this scale are intrinsically more common in this population. A Gaussian albedo distribution, centered at $p=0.05$ and clipped such that $p>0.05$, is consistent with observations for widths $0.058 \leq \sigma_p \leq 0.1$. This estimate does not account for the possibility that binary fraction might vary substantially with primary radius as predicted by Nesvorn\'{y} et al. (2011).

\item All seven systems characterized in this work are widely separated enough to have a substantial probability of disruption if ever subjected to close encounters with Neptune (Parker \& Kavelaars 2010), suggestive of \textit{in situ} formation or a more gentle migration mechanism. Additionally, the collisional lifetimes of these binary systems are short if the size distribution of the Kuiper Belt is steep ($q\sim 3.5$) at small sizes; for the present population of binaries to have survived for the age of the solar system requires that they were never subjected to a period of intense collisional grinding (eg., Petit \& Mousis 2004, Nesvorn\'{y} et al. 2011) and that significant collisions have remained relatively rare in the Kuiper Belt over the age of the solar system. We explore the collisional evolution of these binary systems and further constrain the collisional environment of the present-day Kuiper Belt in an upcoming paper.

\item A confluence of results --- \textit{(a)} gravitational collapse producing similar binary properties as are observed, \textit{(b)} the evidence that the cold Classical Kuiper Belt was not subjected to a period of close encounters with Neptune, and \textit{(c)} the mutual orbits of the wide binaries suggesting formation in a very dynamically-cold environment, yet the $L_2 s$ mechanism not dominating the binary formation process --- all suggest that the cold Classical Kuiper Belt formed in situ through rapid accretion of small particles directly into large ($10-100$ km) planetesimals, driven by turbulent concentration of solids and gravitational collapse.

\end{enumerate}

\section{Acknowledgements}

We thank Melissa Graham for pointing out that the Malmquist bias lurks everywhere, and David Nesvorn\'{y} for providing the results from his simulations of binary formation. We also thank the Gemini staff for their technical support of our observing programs. Alex Parker is funded by the NSF-GRFP award DGE-0836694. This research used the facilities of the Canadian Astronomy Data Centre operated by the National Research Council of Canada with the support of the Canadian Space Agency.

\end{document}